\documentclass[twocolumn,showpacs,preprintnumbers,amsmath,amssymb]{revtex4}
\usepackage{graphicx}
\usepackage{dcolumn}
\usepackage{bm}
\newcommand{\f}{\frac}
\newcommand{\bfll}{\begin{flushleft}}
\newcommand{\efll}{\end{flushleft}}
\newcommand{\bt}{\begin{tabular}}
\newcommand{\et}{\end{tabular}}
\newcommand{\bce}{\begin{center}}
\newcommand{\ece}{\end{center}}
\newcommand{\ben}{\begin{enumerate}}
\newcommand{\een}{\end{enumerate}}
\newcommand{\be}{\begin{equation}}
\newcommand{\ee}{\end{equation}}
\newcommand{\eq}{eqnarray}
\newcommand{\lk}{\left(}
\newcommand{\rk}{\right)}
\newcommand{\ltk}{\left\{}
\newcommand{\rtk}{\right\}}
\newcommand{\ldk}{\left[}
\newcommand{\rdk}{\right]}
\newcommand{\la}{\langle}
\newcommand{\ra}{\rangle}
\newcommand{\nn}{\nonumber \\}
\newcommand{\mH}{\mathcal{H}}
\newcommand{\smn}{\sigma_{\mu\nu}}
\newcommand{\sxy}{\sigma_{xy}}
\newcommand{\sxx}{\sigma_{xx}}
\newcommand{\w}{\omega}
\newcommand{\g}{\gamma}
\newcommand{\ld}{\lambda}
\newcommand{\eps}{\varepsilon}

\newcommand{\alpk}{\alpha_{\bm{k}}}
\newcommand{\bk}{\bm{k}}

\newcommand{\xk}{\xi^{x}_{\bm{k}}}

\newcommand{\yk}{\xi^{y}_{\bm{k}}}
\newcommand{\xy}{\xi^{xy}_{\bk}}
\newcommand{\enp}{E^{+}_{k}}
\newcommand{\enm}{E^{-}_{k}}
\newcommand{\enpm}{E^{+}_{k}-E^{-}_{k}}

\newcommand{\gsim}{\hspace{0.3em}\raisebox{0.4ex}{$>$}\hspace{-0.75em}\raisebox{-.7ex}{$\sim$}\hspace{0.3em}}
\newcommand{\lsim}{\hspace{0.3em}\raisebox{0.4ex}{$<$}\hspace{-0.75em}\raisebox{-.7ex}{$\sim$}\hspace{0.3em}}

\begin{document}

\title{Theory of AC Anomalous Hall Conductivity in $d$-electron systems}

\author{T. TANAKA}
\author{H. KONTANI}
\affiliation{ Department of Physics, Nagoya University, Furo-cho, Nagoya 464-8602, Japan }%

\date{\today}

\begin{abstract}
To elucidate the intrinsic nature of anomalous Hall effect (AHE) in $d$-electron systems, we study the AC anomalous Hall conductivity (AHC) in a tight-binding model with ($d_{xz},d_{yz}$)-orbitals. We drive an analytical expression for the AC AHC $\sxy(\w)$, which is valid for finite quasiparticle damping rate $\g$=$\hbar/2\tau$, and find that the AC AHC is strongly dependent on $\g$. When $\g=+0$, the AC AHC shows a spiky peak at finite energy $\Delta$ that originates from the interband particle-hole excitation, where $\Delta$ represents the minimum band-splitting measured from the Fermi level. In contrast, we find that this spiky peak is quickly suppressed when $\g$ is finite. 
By using a realistic value of $\g(\w)$ at $\w=\Delta/2$ in $d$-electron systems, the spiky peak is considerably suppressed. In the present model, the obtained results also represent the AC spin Hall conductivity in a paramagnetic state.
\end{abstract}

\pacs{72.10.-d, 72.80.Ga, 72.25.Ba, 72.25.-b}

\maketitle

%
%

\section{\label{sec:level1} INTRODUCTION }

In usual metals, the ordinary Hall effect due to Lorentz force is widely observed.
In this case, the Hall resistivity is proportional to the applied magnetic field.
In ferromagnets, in addition, the Hall resistivity that is proportional to the magnetization is observed, which is called the anomalous Hall effect (AHE).
Therefore, the conventional expression for the Hall resistivity, $\rho_H$, is given by $\rho_H = R^0_{H}B + 4\pi R^a_H M$, where $R^0_H$ and $R^a_H$ are the ordinary and anomalous Hall coefficients, $B$ is the magnetic field, and $M$ is the magnetization. 
The mechanism of the AHE has been intensively studied for a long time.

The study by Karplus and Luttinger (KL) \cite{KL} in 1954 was the first theoretical 
approach to the AHE, and it was refined by Luttinger \cite{Luttinger} in 1958.
They pointed out that the anomalous Hall conductivity (AHC) $\sxy^{a}(=R^a_H 
M/\rho^2)$ is finite and dissipation-less ($\sxy^a\propto \rho^0$) when $M \neq 0$ 
in multiband systems with the spin-orbit interaction (SOI). This KL-term is 
called the ``intrinsic AHE" because it is due to the interband particle-hole excitation, 
and it exists even in systems without impurities. 
On the other hand, alternative mechanism due to impurities was proposed as the ``extrinsic AHE".
Smit have shown that extrinsic AHC due to the impurity skew scattering follows $\sxy^{sk}\propto \rho^{-1}$ \cite{Smit}. Later, Berger found that another extrinsic mechanism, the side jump mechanism, gives $\sxy^{sj} \propto \rho^{-2}$ \cite{Berger}.
Recently, AHE due to skew scattering and side jump mechanism has been studied based on linear response theory and semiclassical Boltzmann equation approach \cite{Bruno, Sinitsyn07,Sinitsyn-review}.

After KL, intrinsic AHE has been studied by several specific theoretical models \cite{Sinitsyn07, Sinitsyn-review, Fukuyama, Kontani94, Kontani97, Miyazawa, Sundaram, Nagaosa, Fang, Yao, Kontani06}.
Recently, AHE in the Rashba 2D electron systems has been intensively studied \cite
{Culcer,Dugaev,Sinitsyn05, Nunner, Inoue-AHE, Kato}.
In refs. \cite{Nunner, Inoue-AHE, Kato}, they have reported that the AHC 
vanishes in the Rashba model due to the cancellation by the current vertex correction 
(CVC) due to impurities unless the lifetime is spin-dependent.
In graphene system, the appearance of large quantum spin Hall effect has been 
predicted when the Fermi level lies inside the gap \cite{Sinitsyn-graphene, Kane, 
Yao-graphene}.

In general, intrinsic AHC is composed of the ``Fermi surface term" and the ``Fermi sea term" \cite{Streda}. Recently, refs. \cite{Sundaram,Nagaosa} have reported that KL's AHC, which is a part of the ``Fermi sea term", is expressed in terms of the ``Berry curvature" in case of $\g=+0$, where $\g$ is the quasiparticle damping rate. 
In several simple models \cite{Kontani06, Nagaosa, Sundaram, Sinitsyn-graphene}, the Berry curvature term gives the correct AHC since another Fermi sea term exactly cancels with the Fermi surface term.
However, it presents erroneous result in the high resistive regime since the cancellation of other terms becomes imperfect.
\cite{Kontani06,Kontani-Pt,Tanaka-4d5d}.


AC transport phenomena have been attracting much interest for a long time since 
they can provide detailed and decisive information on the magnetic properties and the 
electronic structure of magnetic materials. For example, the AC AHE have been studied
as the well-known magneto-optical effect (MOE).
In transition metal ferromagnets,
theoretical studies of MOE had been reported
in refs. \cite{Wang, Ebert, Oppeneer}. Therein, the overall behavior of the experimental
AC AHCs at high freqencies are reproduced in many transition metals. 
Moreover, in high-$T_C$ superconductors (HTSCs), AC Hall effects 
have been intensively studied by Drew's group. They have found that the Hall 
coefficient $R_{H}(\w)$ in HTSCs shows a prominent $\w$-dependence \cite
{Drew-Kontani,Drew-YBCO04,Drew-YBCO02,Drew-YBCO00,Drew-YBCO96,Drew-LSCO,
Drew-PCCO}, which can not be reproduced by the extended Drude form.
This anomalous AC transport phenomena can be explained by the fluctuation-exchange 
(FLEX) approximation by considering the CVC \cite
{Kontani-RHletter,Kontani-RHfull,Kontani-review}.
These studies on AC transport phenomena showed the significance of the strong 
antiferromagnetic fluctuation in HTSCs.
In similar, AC AHE at low frequencies will be a fruitful study to understand the intrinsic 
nature of AHE.

Recently, AC AHEs in SrRuO$_3$ and Fe were 
calculated based on the LDA band calculations \cite{Fang,Yao}. According to the $\it{ab \ initio}$
calculation \cite{Niu}, the AC spin Hall effect (SHE) for $p$-type semicondutors was also studied. In ref. \cite{Fang}, they have predicted that AC AHC has a sharp and spiky peak at low frequencies that originates from the interband transition under the assumption that $\g=+0$.
However, in metallic systems, the damping rate $\g(\w)$ should be finite for $\w\neq 0$ even at zero temperature because of inelastic scattering: In a Fermi Liquid, the quasiparticle damping rate is given as $\g(\w)\propto \ltk(\pi T)^2 + \w^2\rtk$, where $T$ represents temperature. 
Therefore, a reliable calculation on the AC AHC for finite $\g$ is highly required. 
 

The aim of this paper is to obtain the reliable expression for the AC AHC for finite $\g$ based on the 
$(d_{xz},d_{yz})$- orbital tight-binding model. For this purpose, we derive an analytical 
expression for the AC AHC including both the Fermi surface and Fermi sea terms 
based on the linear response theory. This expression is valid  for finite $\g$ if the CVC 
is negligible. Using this expression, the AC AHC at low frequencies, where $\w\lsim 4000$K, is studied in detail.
We find that the intrinsic AC AHC shows a non Drude-like behavior. In the case of 
$\g=+0$, it has a sharp and spiky peak at $\w=\Delta$. Here, $\Delta$ is the minimum 
band-splitting measured from the Fermi level, and $\Delta \gsim 1000$K in usual 
transition metals. However, this spiky peak is 
easily suppressed by finite $\g$. By using a realistic value of $\g(\w)$ at 
$\w=\Delta/2$ in usual metals, the spiky peak is almost smeared out. We also find that 
the overall behavior of the AC AHC is reproduced well by the Fermi surface term, 
whereas the Berry curvature term reproduces the correct AC AHC only when $\g\ll 
\Delta$.
Since this condition is not satisfied in usual metals for $\w\sim \Delta$, the Berry 
curvature term gives erroneous AC AHC in the real metallic systems.  

Now, we explain that AC AHE at low frequencies can provide important information 
on the mechanism of the AHE.
For a long time, the origin of AHE (intrinsic or extrinsic) in real systems has been unsettled problem.
In clean heavy fermion systems, $\sxy^a$ is independent of $\rho$ (i.e. $R_H\propto \rho^2$)
sufficiently below the coherent temperature $T_0$, whereas $\sxy^a\propto \rho^{-2} \  (R_{H}\propto\rho^0)$ above $T_0$
\cite{Namiki,Otop,Sullow,Hiraoka}.
This fact indicates that the intrinsic AHE is dominant in clean samples \cite{Kontani94}. 
Also, recent experiments for several transition metal complexes have
reported that $\sxy^a$ is independent of $\rho$ in the low resistive regime \cite{Ong,Asamitsu}, whereas $\sxy^a$ decreases in proportion to $\rho^{-2}$ in the high resistive regime \cite{Asamitsu}. 
Based on the intirinsic AHE in ($d_{xz},d_{yz}$)-orbital tight-binding model,
Kontani et al. \cite{Kontani06} have explained this experimental 
result in transition metal ferromagnets. 
However, in DC AHE, it is not easy to distinguish experimentally which mechanism (intrinsic or extrinsic)
is dominant since the introduction of randomness, by which both the resistivity $\rho$ and the skew scattering increase, makes the analysis of experimental results difficult.
In contrast, AC AHE may be useful to solve this problem by measuring disorder free samples:
Since the AHC due to the skew-scattering mechanism is proportional to $\g^{-1}$ if elastic scattering is dominant, 
the AC AHC due to the this mechanism shows a sensitive $\w$-dependence like the Drude-type behavior:
$\sxy^{sk}(\w) \propto \lk \g - i\w \rk^{-1}$ for small $\w$. On the other hand, AC AHC due to the intrinsic mechanism is independent of $\w$ for $\w\ll\Delta$, as shown in this paper.
Therefore, AC AHE will be useful to resolve the controversy over the origin of the AHE without necessity to introduce disorders.

Finally, we comment on the AC SHE: SHE is the phenomenon that an applied electric field induces a spin current in a transverse direction.
Recently, refs. \cite{KontaniSHE,Kontani-Pt,Tanaka-4d5d} have found that the huge SHE is ubiquitous in multiorbital $d$-electron systems.
The origin of the huge SHE is the ``effective Aharonov-Bohm (AB) phase" induced by the atomic SOI with the aid of inter-orbital hopping inegrals. 
Since $s_z$-spin current operator $j^z_x$ is given by $ j^z_x=-\f{\hbar}{e}\lk j_{x\uparrow}-j_{x\downarrow} \rk$ in the present model, the relation $\sxy^z = -\f{\hbar}{e}\sxy$ holds. Here, $ j_{x\uparrow}(j_{x\downarrow})$ is the charge current operator for $\uparrow(\downarrow )$-spin, and $\sxy^z\equiv j^z_x/E_y$.
Therefore, interesting $\w$-dependence of AC AHC derived in the present study 
is also expected to be realized in AC spin Hall conductivity (SHC).

%
%

\section{\label{sec:level2} MODEL AND HAMILTONIAN}

In this paper, we study a square lattice tight-binding model with $d_{xz}$- and $d_{yz}$- orbitals, which is a simplified model of a famous triplet superconductor in Sr$_2$RuO$_4$ \cite{Mackenzie}. 
The $(d_{xz},d_{yz})$-orbital tight-binding model is one of the simplest models for studying the AHE in transition metal ferromagnets \cite{Kontani06}.
Using this model, we derive explicit expressions for AC AHC that is valid for finite quasiparticle damping rate. 

To realize the AHE in ferromagnets, the SOI is indispensible. 
In a perfect ferromagnetic metal with $\bm{M}||\hat z$, the atomic SOI $\ld \bm{\hat l}\cdot \bm{\hat s}$ is replaced with $-\ld (M/\mu_{B} )\hat l_z$, where $\ld$ is the coupling constant. The intrinsic AHE is caused by the interband transition of quasiparticles due to the off-diagonal elements of $\hat l_z$ \cite{KL}. In the $d$-electron systems, the matrix element of $\hat l_z$ is finite only for $\langle yz | l_z| zx \rangle= -\langle zx|l_z| yz\rangle = i$ and $\langle xy |l_z | x^2-y^2 \rangle = -\langle x^2-y^2 |l_z | xy \rangle =2i $. Note that $| zx \rangle$ and $| yz\rangle$ are given by the linear combination of $l_z =\pm 1$, and $| xy \rangle$ and  $| x^2-y^2 \rangle$ are given by the linear combination of $l_z =\pm 2$. 
In pure transition metals, since the energy splitting between $t_{2g}$ and $e_g$ orbitals is smaller than the bandwidth, the interband transition of the quasiparticle between $d_{xy}$-orbital (in $t_{2g}$) and $d_{x^2-y^2}$-orbital (in $e_g$) should be taken into account correctly. In fact, we have found that the SHE is mainly caused by $(d_{xy},d_{x^2-y^2})$-orbital in pure transition metals \cite{Tanaka-4d5d}.
On the other hand, in ruthenates, since the energy splitting between $t_{2g}$ and $e_g$ orbitals are large,
AHE is maily caused by $(d_{xz},d_{yz})$-orbital.
In fact, $(d_{xz},d_{yz})$-orbital tight-binding model can explain important experimental fact as reported in ref. \cite{Asamitsu}.
We note that the mechanisms of large AHEs arises from $(d_{xz},d_{yz})$-orbitals and $(d_{xy},d_{x^2-y^2})$-orbitals are the same: The ``effective AB phase" factor of conduction electrons due to $d$-atomic angular momentum with the aid of the atomic SOI and the inter-orbital hoppings \cite{KontaniSHE,Kontani-Pt,Tanaka-4d5d}.

\begin{figure}[!htb]
\includegraphics[width=.9\linewidth]{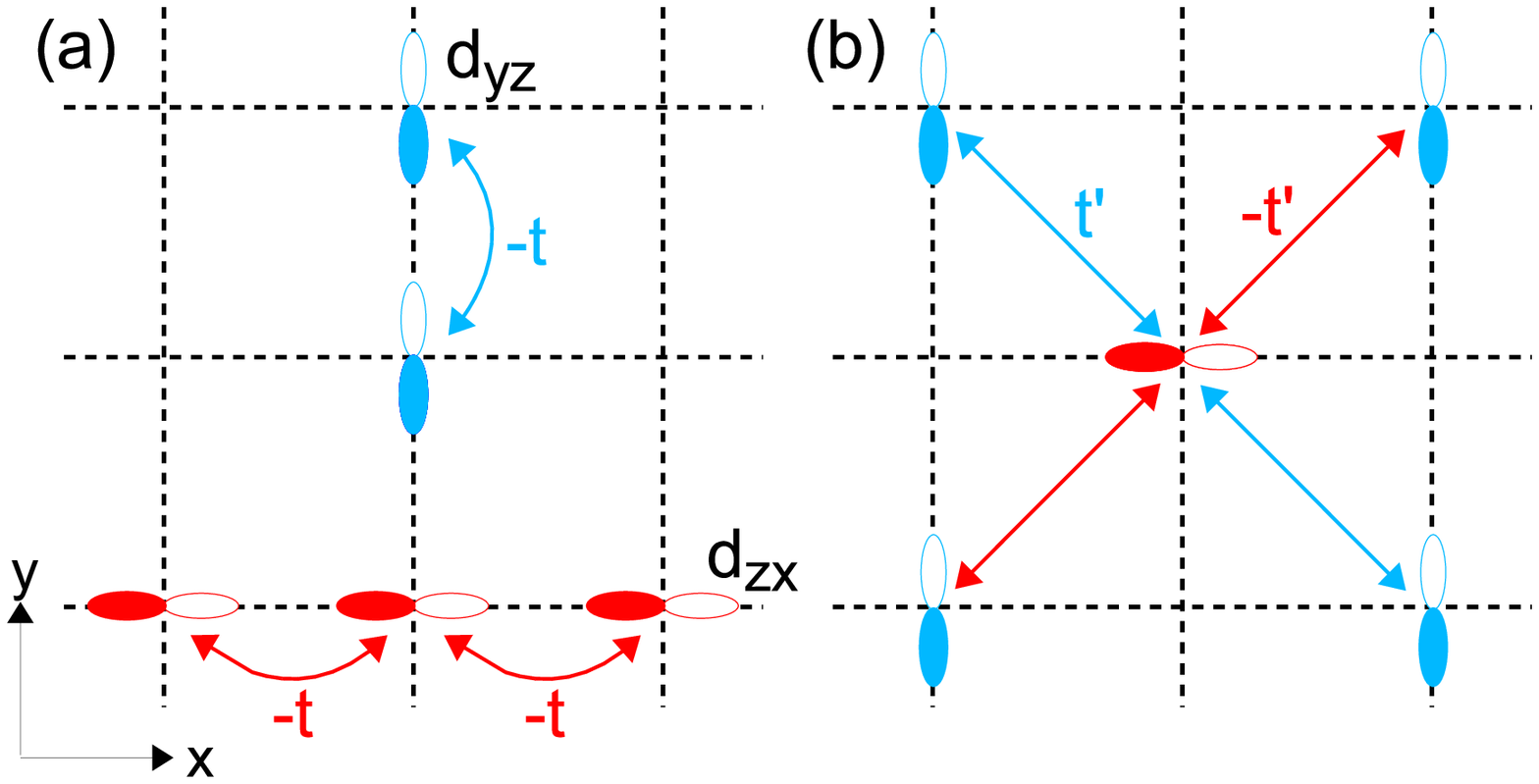}
\caption{\label{fig:d-hop} (a) Hopping integrals between the same orbitals. (b) Hopping integrals between the different orbitals, which is given by the nearest neighbor hopping.
Note that the sign of the interorbital hopping integrals change by $\pi/2$ rotaion. This fact gives rise to the anomalous velocity.}
\end{figure}

Here, we represent the creation operator of an electron on $d_{xz}$-($d_{yz}$-) orbital as $\hat c^{x\dagger}_{\bk}(\hat c^{y\dagger}_{\bk})$. The Hamiltonian without
SOI is given by $H^0=\sum_{\bk}\hat c^{\dagger}_{\bk} \hat h^0_{\bk} \hat c_{\bk}$, where \cite{Kontani06}
\begin{align}
\hat h^0_{\bk}=
\left(
\begin{array}{cc}
\xi^x_{\bk} & \xi^{xy}_{\bk} \\
\xi^{xy}_{\bk} & \xi^y_{\bk} 
\end{array}
\right),  
\end{align}
and $\hat c^{\dagger}_{\bk}=(\hat c^{x\dagger}_{\bk},\hat c^{y\dagger}_{\bk})$. $\xi^x_{\bk}=-2t\cos k_x, \ \xi^y_{\bk}=-2t\cos k_y$ 
and $\xi^{xy}_{\bk}=4t'\sin k_x\sin k_y$, where
$-t$ and $\pm t'$ are the hopping integrals between nearest-neighbors and nextnearest-neighbors, respectively.
They are shown in Fig. \ref{fig:d-hop}.

In the present model, the velocity matrix $\hat v_{\mu}=\partial \hat h^{0}_{\bm{k}}/\partial k_{\mu}$
($\mu=x,y$) is given by
\begin{align}
\hat v_x&=
\left(
\begin{array}{cc}
2t\sin k_x & 4t' \cos k_x\sin k_y \\
4t' \cos k_x\sin k_y & 0
\end{array}
\right),  \\
\hat v_y&=
\left(
\begin{array}{cc}
0 & 4t'\sin k_x\cos k_y \\
4t'\sin k_x\cos k_y & 2t\sin k_y
\end{array}
\right).
\end{align}

We should stress that the off-diagonal elements of $\hat v_x$, $v^{xy}_x=v^{yx}_x$, are odd-functions of $k_y$. In the same way, $v^{xy}_{y}=v^{yx}_y$ 
are odd-functions of $k_x$. They are called the ``anomalos velocity". In later sections, we will see that AHC $\sxy^a$ is proportional to $\ld \la v^{xx}_xv^{xy}_y\ra$.
This implies that sizable AHC is caused by anomalous velocity with the aid of atomic SOI.
Consequently, atomic d-orbitals degree of freedom gives rise to the huge AHC in transition metal ferromagnets.

Next, the atomic SOI, which is indispensible for AHE, is given by
$\mH^{\ld}=\sum_{\bk}\hat c^{\dagger}_{\bk}\hat h^{\ld}\hat c_{\bk}$.
Here, $\hat h^{\ld}$ in the present bases is given by \cite{Kontani06}
\begin{align}
\hat h^{\ld}=-\text{sgn}(s_z)\ld\hat \tau_y,
\end{align}
where $\hat \tau_y$ represents the Pauli matrix for the orbital space. Hereafter, we put $\mu_B=1$ for the 
simpilcity of calculation.

\begin{figure}[!htb]
\includegraphics[width=.85\linewidth]{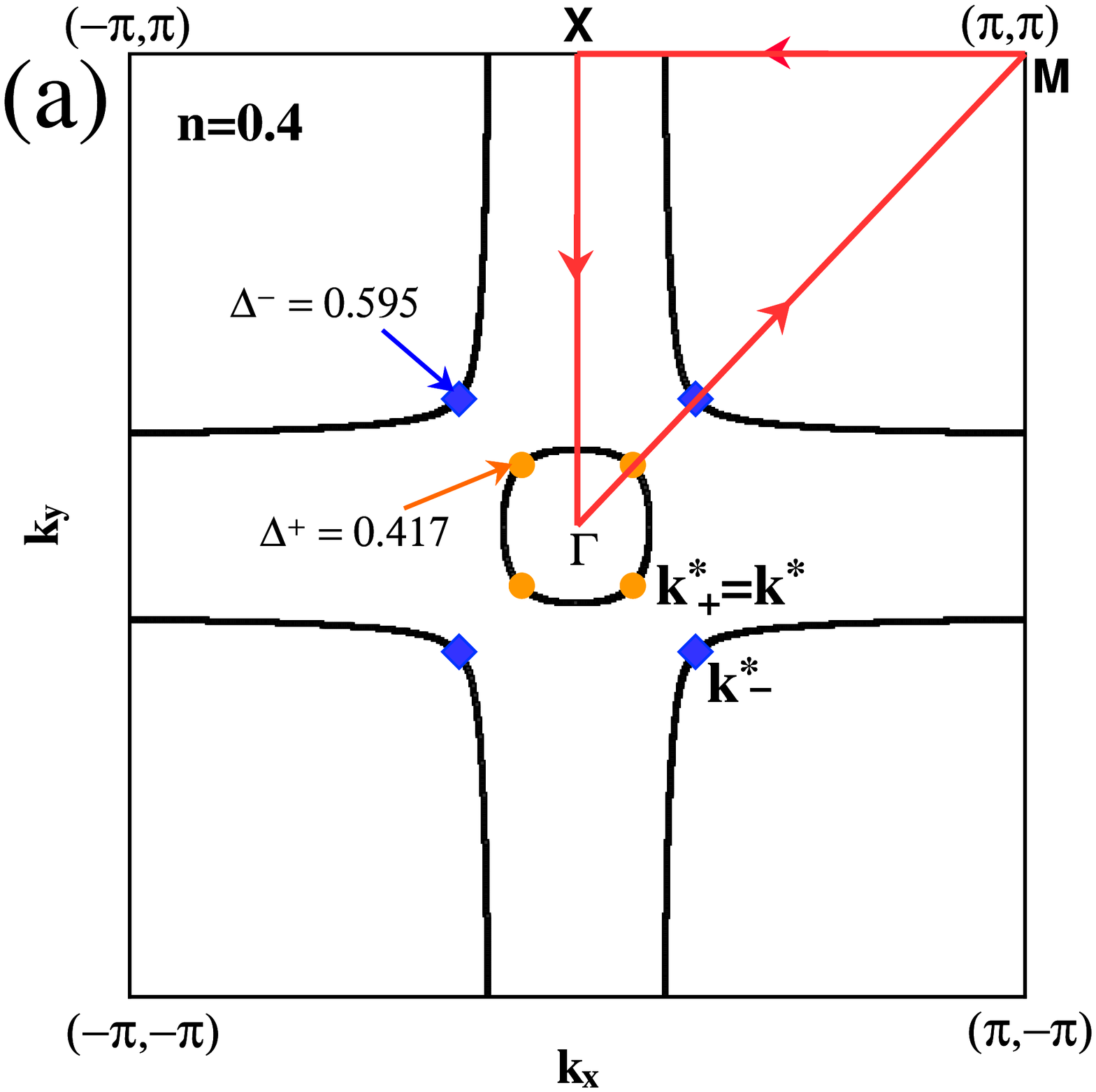}
\includegraphics[width=0.9\linewidth ,height=0.7\linewidth]{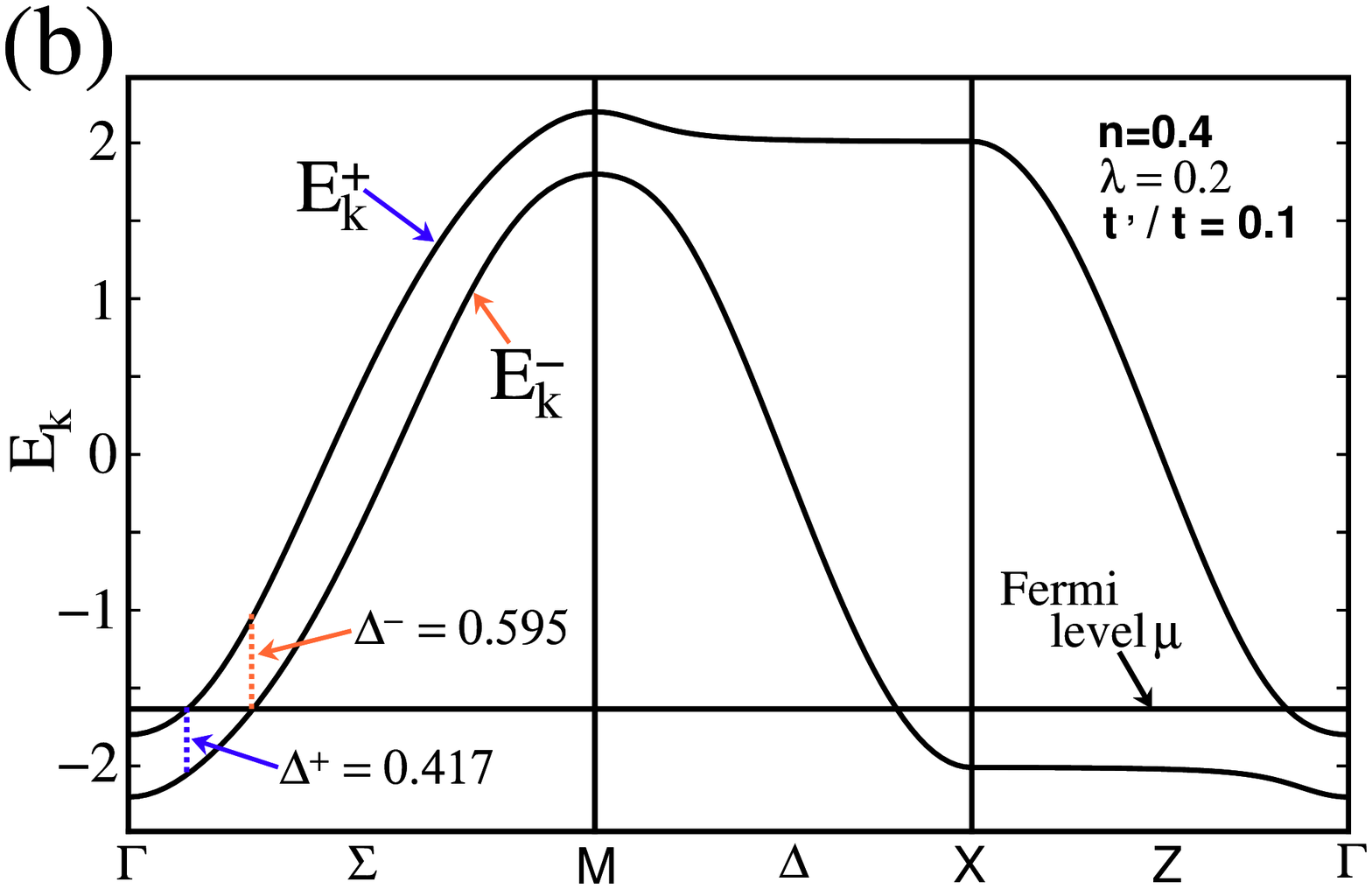}
\caption{\label{FS} (a) Fermi surface for $n=0.4$. Here we put $(t,t')=(1,0.1)$, which corresponds to $\text{Sr}_2\text{Ru0}_4$. Here, $\Gamma=(0,0)$, $M=(\pi,\pi)$, and $X=(0,\pi)$. (b) Band structure in the present model. Here, we set $n$=0.4, $\lambda$=0.2 and $t/t'=0.1$.
}
\end{figure}

The Green function in the presence of atomic SOI is given by $\hat G_{\bk}(\w)
=\lk \w+\mu-\hat h^0_{\bk}-\hat h^{\ld}\rk^{-1}$, which is expressed as follows in the present model \cite{Kontani06}:
\begin{align}
\hat G(\w)&= \left(
\begin{array}{cc}
G_{xx}(\w) & G_{xy}(\w) \\
G_{yx}(\w) & G_{yy}(\w)
\end{array}
\right) \\
&=\f{1}{d(\w)}\left(
\begin{array}{cc}
\w+\mu-\yk & \alpk \\
\alpk^{\ast} & \w+\mu-\xk
\end{array}
\right), \label{eq:Gfunc}
\end{align}
where $\alpk=\xy-i\ld \text{sgn}(s_z)$ and $d(\w)=(\w+\mu-\xk)(\w+\mu-\yk)-|\alpk|^2$, which is expressed as
\begin{align}
d(\w)=&(\w+\mu-\enp)(\w+\mu-\enm),  \\
E^{\pm}_{\bk}=&\f{1}{2}\lk \xi^{x}_{\bk}+\xi^y_{\bk} \pm \sqrt{(\xi^x_{\bk}-\xi^y_{\bk})^2+4|\alpha_{\bk}|^2} \rk. 
\end{align}
Here, $E^{\pm}_{\bk}$ represents the quasiparticle dispersion. Figures \ref{FS} (a) and (b) show the Fermi surface and the band structure obtained in the present model for $(t,t')$ = $(1,0.1)$, which is realized in $\text{Sr}_2\text{RuO}_4$ \cite{Nomura}. The electron density per spin, $n$, is set as 0.4. The energy splitting $\Delta^{\pm}$ represents the minimum band-splitting $(|E^+_{\bk}-E^-_{\bk}|)$ measured from the Fermi surface of $E^{\pm}_{\bk}$-band. 
In Fig. \ref{FS}, $\Delta^{\pm}=|E^+_{\bk}-E^{-}_{\bk}|$ at $k^{\ast}_{\pm}$, and
$k^{\ast}$ represents the position of the minimum band-splitting $\Delta\equiv\text{min}\ltk \Delta^+, \ \Delta^- \rtk$.

Here, we consider the quasiparticle damping rate $\hat \Gamma(\w)$. Microscopically, it is 
given by the imaginary part of the self-energy, $\hat \Sigma_{\bk}(\w) $. For simplicity, we assume that $\hat \Gamma(\w)$ is diagonal with respect to orbitals,
and is independent of the momentum: $\Gamma_{\alpha\beta}(\w)=\g(\w)\delta_{\alpha\beta}$, where $\alpha$ and $\beta$ are orbital indices. This assumption is justified in the case of the elastic scattering due to local impurities, since the local
Green function $g_{\alpha\beta}=\sum_{\bk} G_{\alpha\beta}$ is small for $\alpha\neq\beta$ in the case of $\ld \ll W_{band}$, where $W_{band}$ is the bandwidth \cite{Kontani06}.
This fact is also justified in the case of inelastic scattering due to on-site Coulomb interaction
in the dynamical mean field approximation (DMFA), where the self-energy is composed of local Green functions.
Using eq. (\ref{eq:Gfunc}), the retarded and advanced Green functions are given by
\begin{eqnarray}
G^R_{\alpha\beta'}(\w)=G_{\alpha\beta'}(\w+i\g(\w)),  \\
G^A_{\alpha\beta'}(\w)=G_{\alpha\beta'}(\w-i\g(\w)),
\end{eqnarray}
respectively.
Hereafter, we assume that the renormalization factor $\hat z_{\bk}\equiv \left( 1- \partial\hat \Sigma/\partial \w \right)^{-1}=1$ since it cancels out in the final expression of the AHE \cite{Kontani06}. 

%
%

\section{\label{sec:level3} AC HALL CONDUCTIVITY}

In this section, we derive an analytical expressions for the AC Hall conductivity $\sxy(\w)$ and the AC 
longitudinal conductivity $\sxx(\w)$ based on the linear response theory by dropping all the CVC.
The other approach to the AHE, which is the semiclassical Boltzmann equation approach, is also useful \cite{Sundaram,Sinitsyn07,Sinitsyn-review}: In ref. \cite{Sinitsyn07}, the equivalence of these two methods has been shown in the 2D Dirac-band graphene system.

Until section \ref{sec:level5}, we assume that the elastic scattering due to the impurity potential 
is dominant over the inelastic scattering due to electron-electron interaction. 
In this case, the quasiparticle damping rate is given by $\displaystyle \g_{imp}(\w)= \f{n_{imp}\pi I^2N(\w)}{1-\pi IN(\w)}$, where $N(\w)$ is the density of states per orbital, $n_{imp}$ is the impurity concentration, and $I$ is the impurity potential. Since the $\w$-dependence of $N(\w)$ is small in the present model, we omit the $\w$-dependence of $\g_{imp}$.
In section \ref{sec:level5}, we study the AC Hall conductivity in systems where inelastic scattering is 
dominant. In this case, we have to take account of $\w$ dependence of the damping rate.

Here, we comment on the CVC due to the local impurity potential.
In the Born approximation, the lowest order CVC is given by $\displaystyle \Delta \hat J_{\mu}=n_{imp}I^2 \frac{1}{N} \sum_{\bk} \hat G^R \hat J_{\mu} \hat G^A$. 
In the $d$-orbital tight-binding models, $\hat G$ has an even parity with respect to $\bk \leftrightarrow -\bk$. Therefore, $\Delta\hat J_{\mu}=0$ since $(\partial/\partial k_{\mu})\hat G =\hat G \hat J_{\mu} \hat G$ is an odd function \cite{Kontani06,KontaniSHE,Tanaka-4d5d}.
In contrast, the CVC plays an essential role in the Rashba models \cite{Nunner,Inoue-AHE,Kato}.

\begin{figure}[!htb]
\includegraphics[width=.65\linewidth]{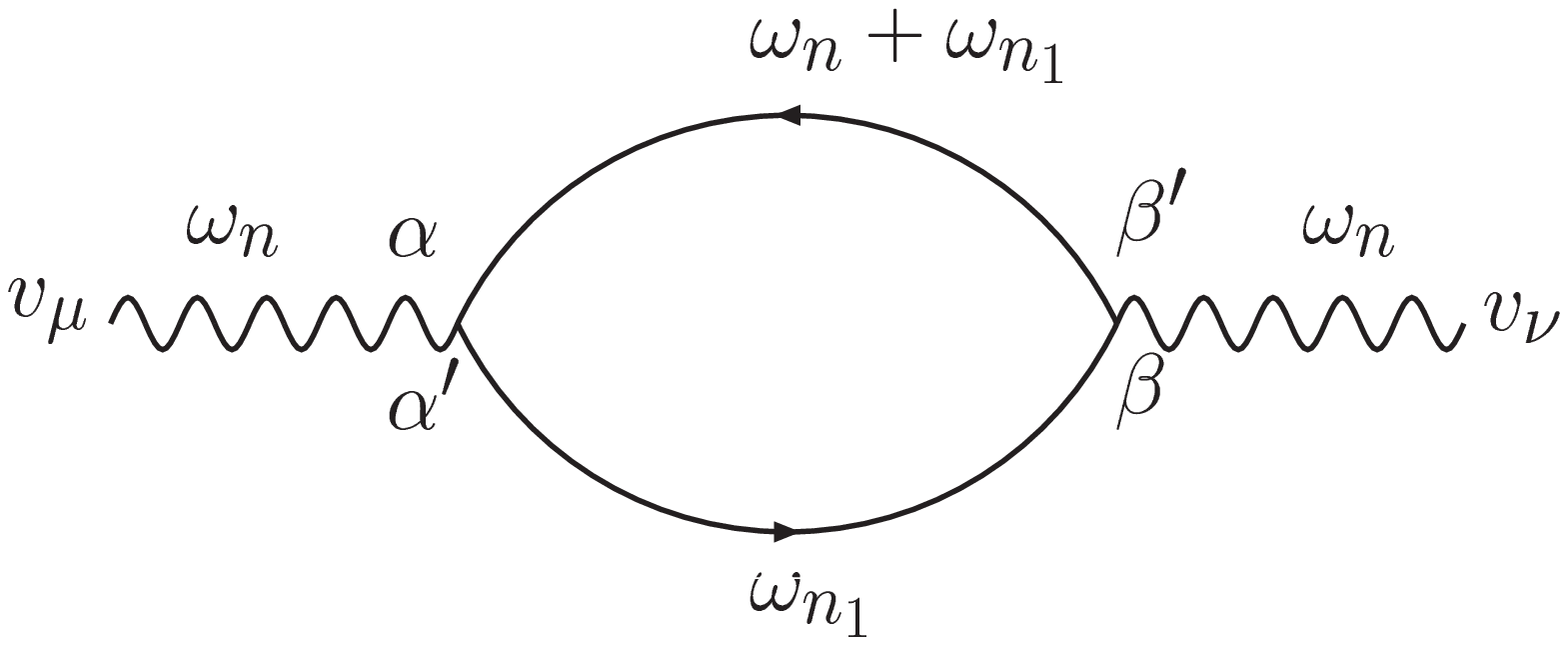}
\caption{\label{fig:diagram} Diagrammatic expression for $\sxy(\w_n)$}
\end{figure}

According to linear response theory, the AC Hall conductivity is given by
\begin{align}
\smn(\w)=\f{1}{i\w}\ldk K^{R}_{\mu\nu}(\w) -K^{R}_{\mu\nu}(0) \rdk, \label{eq:sxy}
\end{align}
where $K^{R}_{\mu\nu}(\w)$ is the retarded correlation function, which is given by the analytic continuations of the following thermal Green function:
\begin{align}
K_{\mu\nu}(\w_n)=e^2\sum_{\bk,\bk',\alpha,\alpha',\beta,\beta'} v^{\alpha'\alpha}_{\bk\mu}v^{\beta'\beta}_{\bk'\nu} Q_{\bk\bk'}(\w_n),
\end{align}
where
\begin{align}
v_{\bk,\mu}=\f{\partial \varepsilon_{\bk}}{\partial k_{\mu}},
\end{align}
and
\begin{align}
Q_{\bk\bk'}(\w_n)=\int^{\beta}_{0}d\tau e^{i\w_n \tau} \la T_{\tau}\lk a^{\dagger}_{\bk\alpha'}(\tau)a_{\bk\alpha}(\tau)a^{\dagger}_{\bk'\beta'}(0)a_{\bk'\beta}(0) \rk \ra.
\end{align} 
If we drop the CVC, $Q_{\bk,\bk'}$ is given by
\begin{align}
Q_{\bk\bk'}(\w_n)=-\delta_{\bk\bk'}\beta^{-1}\sum_{n_1}G_{\beta\alpha'}(\w_{n_1})G_{\alpha'\beta}(\w_n+\w_{n_1}), \label{K0}
\end{align}
The diagrammatic expression of $\smn(\w_n)$ is shown in Fig. \ref{fig:diagram}.

\begin{widetext}
Performing the analytic continuation carefully, the expression for the AC Hall conductivity $\sxy(\w)$ is given by
%
\begin{\eq}
\sxy(\w)&=&\sum_{\bk,\alpha',\alpha,\beta',\beta}\int \f{d\eps}{2\pi}\f{-1}{\w}v^{\alpha'\alpha}_{x}v^{\beta'\beta}_{y} \nn
         && \times \ldk \ltk f(\eps+\w/2) - f(\eps-\w/2) \rtk G^{R}_{\alpha\beta'}(\eps+\w/2) G^{A}_{\beta\alpha'}(\eps-\w/2) \right. \nn
&& \left. + f(\eps+\w/2) G^{R}_{\alpha\beta'}(\eps+\w/2)  G^{R}_{\beta\alpha'}(\eps-\w/2)  - f(\eps-\w/2) G^{A}_{\alpha\beta'}(\eps+\w/2)  G^{A}_{\beta\alpha'}(\eps-\w/2) \right]  \label{basesxy}
\end{\eq} 
where $f(\eps)$ is the Fermi distribution fuction. 
Here, we divided $\sxy(\w)$ into two terms, where the first term in the square bracket corresponds to the Fermi surface term ($I$), and the second and third terms correspond to the Fermi sea term ($II$). 
Since $I$ term consist of $G^RG^A$, whereas $II$ term consist of $G^R G^R$ 
and $G^A G^A$, the division of $\sxy(\w)$ into these two terms is unique.
Hereafter, we drop the factor $e^2/\hbar$ in $\smn(\w)$ to simplify expressions.
Note that $K^R_{xy}(0)=0$ since it is proportional to $G^RG^R-G^AG^A$ \cite{Kontani06}.

Now, we first take the summation over $\alpha', \alpha, \beta', \beta$ in eq. (\ref{basesxy}). 
In the present model, the terms $(\alpha',\alpha,\beta',\beta)=(x,x,x,y),(x,x,y,x),(x,y,x,x)$ and $(y,x,x,x)$ remain finite after $\bk$-summation, just like the calculation of DC AHC in ref. \cite{Kontani06}. Considering the square lattice symmetry of the present model, we obtain the following expression: 
%
\begin{\eq}
\sxy(\w)=\sum_{\bk}&&\f{i\ld}{\pi\w}\int d\eps  \ v^{xx}_xv^{xy}_y  \nn
                 \times &&\ldk \ltk f(\eps+\w/2) - f(\eps-\w/2) \rtk 
\f{\w+2i\g}{d^R(\eps+\w/2)d^A(\eps-\w/2)} \right. \nn 
&& \left.+ f(\eps -\w/2) \f{\w}{d^R(\eps+\w/2)d^R(\eps-\w/2)}
-f(\eps+\w/2) \f{\w}{d^A(\eps+\w/2)d^A(\eps-\w/2)} \rdk 
\end{\eq}
This integration by $\eps$ can be calculated analytically, and the final result for the AC Hall conductivity $\sigma_{xy}(\omega)$ at $T$=0 is given by
\begin{align}
\sxy(\w)&= \sxy^{I}(\w) + \sxy^{IIa}(\w) + \sxy^{IIb}(\w), \label{eq:sxyw} \\
\sxy^{I}(\w)&=\f{i\ld}{\pi\w}\sum_{k}v^{xx}_x v^{xy}_y\f{1}{\enpm} \nn
             &\times\ldk \f{1}{\enpm-\w-2i\g}\ltk\ln\lk\f{\mu+\w-\enp+i\g}{\mu-\enm-i\g}\rk -\ln\lk\f{\mu-\enp+i\g}{\mu-\w-\enp-i\g}\rk \rtk \right.   \nn
             & \left.-\f{1}{\enpm+\w+2i\g}\ltk\ln\lk\f{\mu-\enp-i\g}{\mu+\w-\enm+i\g}\rk -\ln\lk\f{\mu-\w-\enp-i\g}{\mu-\enm+i\g}\rk \rtk \rdk, \label{eq:sxyIw}  \\
\sxy^{IIb}(\w)&=\f{2i\ld}{\pi\w}\sum_{k}v^{xx}_x v^{xy}_y\f{1}{\enpm} \nn
                &\times\ldk \f{1}{\enpm+\w}\ltk \ln\lk \mu-\enp+i\g \rk+\ln\lk \mu-\enm-i\g \rk \rtk \right. \nn
               & \left.+\f{1}{\enpm-\w}\ltk\ln\lk \mu-\enm+i\g \rk+\ln\lk \mu-\enp-i\g \rk\rtk  \rdk,  \label{eq:sxyIIbw} \\
\sxy^{IIa}(\w)&=\f{-i\ld}{\pi\w}\sum_{k}v^{xx}_x v^{xy}_y\f{1}{\enpm} \nn
         &\times\ldk \f{1}{\enpm+\w}\ltk \ln\lk\mu+\w-\enm+i\g \rk+\ln\lk \mu-\w-\enp-i\g \rk \rtk \right. \nn
         & \left. +\f{1}{\enpm-\w}\ltk \ln \lk \mu+\w-\enp+i\g \rk+\ln\lk \mu-\w-\enm-i\g \rk\rtk  \rdk \nn 
&-\f{1}{2} \times [\text{eq}. (\ref{eq:sxyIIbw})], \label{eq:sxyIIaw}             
\end{align}
If we put $\w=0$, the obtained expression reproduces the DC AHC given in ref. \cite{Kontani06}.
%
In the above equations, we divided the Fermi sea term into two terms, which are $\sxy^{IIa}(\w)$ and $\sxy^{IIb}(\w)$: $\sxy^{IIb}$ is called the Berry curvature term \cite{Nagaosa, Sundaram,Niu,NagaosaReview}. 
Here, we analyze eqs. (\ref{eq:sxyIw}), (\ref{eq:sxyIIbw}) and (\ref{eq:sxyIIaw}) in the clean limit $n_{imp}\rightarrow 0$ ($\g\rightarrow 0$), and 
show that the real part of $\sxy^{IIb}(\w)$ in eq. (\ref{eq:sxyIIbw}) corresponds to the expression which are frequently used to calculate $\sxy(\w)$.
%
%
In the band-diagonal representation $(E^{\alpha}_{\bk};\alpha=\pm )$, we can show that the off diagonal velocity is given by \cite{Kontani06}
\begin{align}
v^{\alpha \bar\alpha}_{x}v^{\bar\alpha \alpha}_{y}=&-i\lambda\f{v^{xx}_{x}v^{xy}_{y}+v^{xy}_{x}v^{yy}_{y}}{E^{\alpha}_{\bk}-E^{\bar\alpha}_{\bk}} \nn
&+(\text{terms which vanish with $\bk$-summation.}).
\end{align}
Since $\text{Im} \ldk \ln(x)\pm i\g \rdk= \mp \pi\theta(x)$ for $\g \rightarrow +0$, eq. (\ref{eq:sxyIIbw}) is rewritten as follows in the band-diagonal representation:
\begin{align}
\sxy^{IIb}(\w)=-i\sum_{\bk,m\neq n}\f{\la m|\hat v_x |n\ra \la n|\hat v_y  |m\ra }{\w}\f{f(E^{n}_{\bk})-f(E^{m}_{\bk})}{\w +E^{n}_{\bk}-E^{m}_{\bk}+i\delta}, \label{eq:BCT}
\end{align}
where $m,n$ are the band indices, and an infinitesimal imaginary part $i\delta$ is included to maintain the analytic property of the Hall conductivity.
The obtained expression of $\sxy^{IIb}(\w)$ in eq. (\ref{eq:BCT}) corresponds to the expression which are used to calculate $\sxy(\w)$ in refs. \cite{Wang, Ebert, Oppeneer, Niu, NagaosaReview}.

In literatures, eq. (\ref{eq:BCT}) has been frequently used since 
$\sxy^{I}(\w)+\sxy^{IIa}=0$ for $\gamma=+0$. However, this condition for $\g$ 
will not be satisfied in the real metallic systems since $\g$ increases with  $\w$ 
due to inelastic scattering. For this reason, $\sxy(\w)\neq \sxy^{IIb}(\w)$ in 
usual metals. Therefore, for reliable calculation, we have to analyze $I, IIa$, and 
$IIb$ terms on the same footing.




Here, we comment on the physical meaning of the three terms: 
$\sxy^{I}$ and $\sxy^{IIa}$ are finite only when the Fermi surface exists, whereas 
$\sxy^{IIb}$ is finite even if the Fermi surface is absent \cite{Kontani06}.
$\sxy^{IIb}$ term is the origin of the quantum Hall effect and 
the quantum spin Hall effect \cite{Sinitsyn-graphene, Kane, 
Yao-graphene, Thouless, Haldane, Murakami}.

According to eq. (\ref{eq:sxy}), the expression for the AC longitudinal conductivity $\sxx(\w)$ is given by
\begin{align}
\sxx(\w)= & \sum_{\bk,\alpha',\alpha,\beta',\beta}\int \f{d\eps}{2\pi}\f{-1}{\w} \ v^{\alpha'\alpha}_xv^{\beta'\beta}_x  \nn
                       &\times\ltk f(\eps-\w/2)G^{R}_{\alpha\beta'}(\eps+\w/2)\lk G^{R}_{\beta\alpha'}(\eps-\w/2)-G^{A}_{\beta\alpha'}(\eps-\w/2)\rk \right. \nn
                       &+ f(\eps+\w/2)\lk G^{R}_{\alpha\beta'}(\eps+\w/2)-G^{A}_{\alpha\beta'}(\eps+\w/2) \rk G^{A}_{\beta\alpha'}(\eps-\w/2) \nn
                       &\left. -f(\eps)\lk G^R_{\alpha\beta'}(\eps)G^R_{\beta\alpha'}(\eps)-G^A_{\alpha\beta'}(\eps)G^A_{\beta\alpha'}(\eps) \rk \rtk. \label{sxxw} 
\end{align}
In the same way as $\sxy(\w)$, we can calculate $\sxx(\w)$ analytically. 
By dropping the terms that vanish after $\bk$-summation, $\sxx(\w)$ is given by, 
\begin{align}
\sxx(\w)=\sum_{\bk}&\f{-1}{2\pi\w}\ldk (v^{xx}_x)^2 A(\w) +4v^{xx}_xv^{xy}_x B(\w) +2(v^{xy}_x)^2 C(\w) \rdk,  \label{eq:sxx}
\end{align}
where
\begin{align}
A(\w)= \ &I_{12}-2(\yk-i\g)I_{11}-\ltk(\w/2)^2-(\yk-i\g)^2\rtk I_{10} \nn
       &-\ldk I_{32}-2(\yk+i\g)I_{31}-\ltk (\w/2)^2-(\yk+i\g)^2\rtk I_{30} \rdk \nn
       &-\ldk I_{22}-2\yk I_{21}-\ltk(\w/2+i\g)^2-(\yk)^2 \rtk I_{20} \rdk \nn
       &-2i \text{Im}\ldk K_2 -2(\yk-i\g)K_1+(\yk-i\g)^2K_0 \rdk, \\
B(\w)= \ &\xi^{xy}_{\bk} I_{11}-\xi^{xy}_{\bk}(\yk-i\g)I_{10}-\ldk \xi^{xy}_{\bk} I_{31} -\xi^{xy}_{\bk}(\yk+i\g)I_{30}\rdk-\ldk \xi^{xy}_{\bk} I_{21}-\xi^{xy}_{\bk}\yk I_{20}\rdk \nn
&- 2i \xi^{xy}_{\bk} \text{Im} \ldk K_1-(\yk-i\g)K_0 \rdk,  \\
C(\w)= \ &I_{12}-(\xk+\yk-2i\g)I_{11}+\ltk (\xi^{xy}_{\bk})^2-\ld^2+(\xk-i\g)(\yk-i\g)-(\w/2)^2 \rtk I_{10} \nn
       &-\ldk I_{32}-(\xk+\yk+2i\g)I_{31}+\ltk (\xi^{xy}_{\bk})^2-\ld^2+(\xk+i\g)(\yk+i\g)-(\w/2)^2 \rtk I_{30} \rdk  \nn
       &-\ldk I_{22}-(\xk+\yk)I_{21}+\ltk (\xi^{xy}_{\bk})^2-\ld^2+\xk\yk-(\w/2+i\g)^2 \rtk I_{20} \rdk \nn
       &-2i \text{Im}\ldk K_2-(\xk+\yk-2i\g)K_1+\ltk (\xi^{xy}_{\bk})^2-\ld^2+(\xk-i\g)(\yk-i\g)\rtk K_0 \rdk, 
\end{align}
and 
\begin{align*}
&I_{1n}= \int^{+\w/2}_{-\infty}d\eps\f{\eps^n}{d^R(\eps+\w/2)d^R(\eps-\w/2)} , \ 
I_{2n}= \int^{+\w/2}_{-\w/2}d\eps\f{\eps^n}{d^R(\eps+\w/2)d^A(\eps-\w/2)}, \nn
&I_{3n}= \int^{-\w/2}_{-\infty}d\eps\f{\eps^n}{d^A(\eps+\w/2)d^A(\eps-\w/2)} , \   
K_n= \int^{0}_{-\infty}d\eps\f{\eps^n}{d^R(\eps)d^R(\eps)}  \  \ \ (n=0,1,2).  
\end{align*}
These integrals can be performed analytically. As a result, we obtain the expression for the AC longitudinal conductivity $\sigma_{xx}(\w)$ at $T$=0.
We can verify analytically that this expression reproduces the longitudinal conductivity $\sxx$ at $\w=0$, which is given in ref. \cite{Kontani06}.
\end{widetext}

%
%

\section{\label{sec:level4} NUMERICAL STUDY}

In this section, we perform the numerical study for both $\sxy(\w)$ and $\sxx(\w)$ at $T=0$, assuming a perfect ferromagnetic state where $n_{\downarrow}=n$ and 
$n_{\uparrow }=0$. In this case, $m_{z}=\mu_{B}n$. Hereafter, we put $\mu_{B}=1$. Also, we put the coupling constant of SOI as $\ld=0.2$.
It corresponds to 800K if we assume t=4000K, which is a realistic value in ruthenates.
The main purpose of this section is to elucidate the frequency ($\w$) dependence and the damping rate ($\g$) dependence of the AC Hall conductivity. 
We perform the $\bk$-summations in eq. (\ref{eq:sxyw}) for $\sxy$ and in eq. (\ref{eq:sxx}) for $\sxx$ numerically, dividing the Brillouin zone into $5000 \times 5000$ meshes. 
By using the obtained results for $\sxy(\w)$ and $\sxx(\w)$, we also present the $\w$- and $\g$-dependences of the Hall coefficient $R_H(\w)=\sxy(\w)/\sxx^2(\w)$ and the Hall angle $\theta_H(\w)=\sxy(\w)/\sxx(\w)$.

The unit of conductivity in this section is $e^2/ha$, where $h$ is the plank constant and $a$ is the unit cell length. If we assume the length of
unit cell $a$ is $4 \text{\AA}$, then $e^2/ha\approx 10^3\Omega^{-1}cm^{-1} $.

\begin{figure}[!htb]
\includegraphics[width=.85\linewidth]{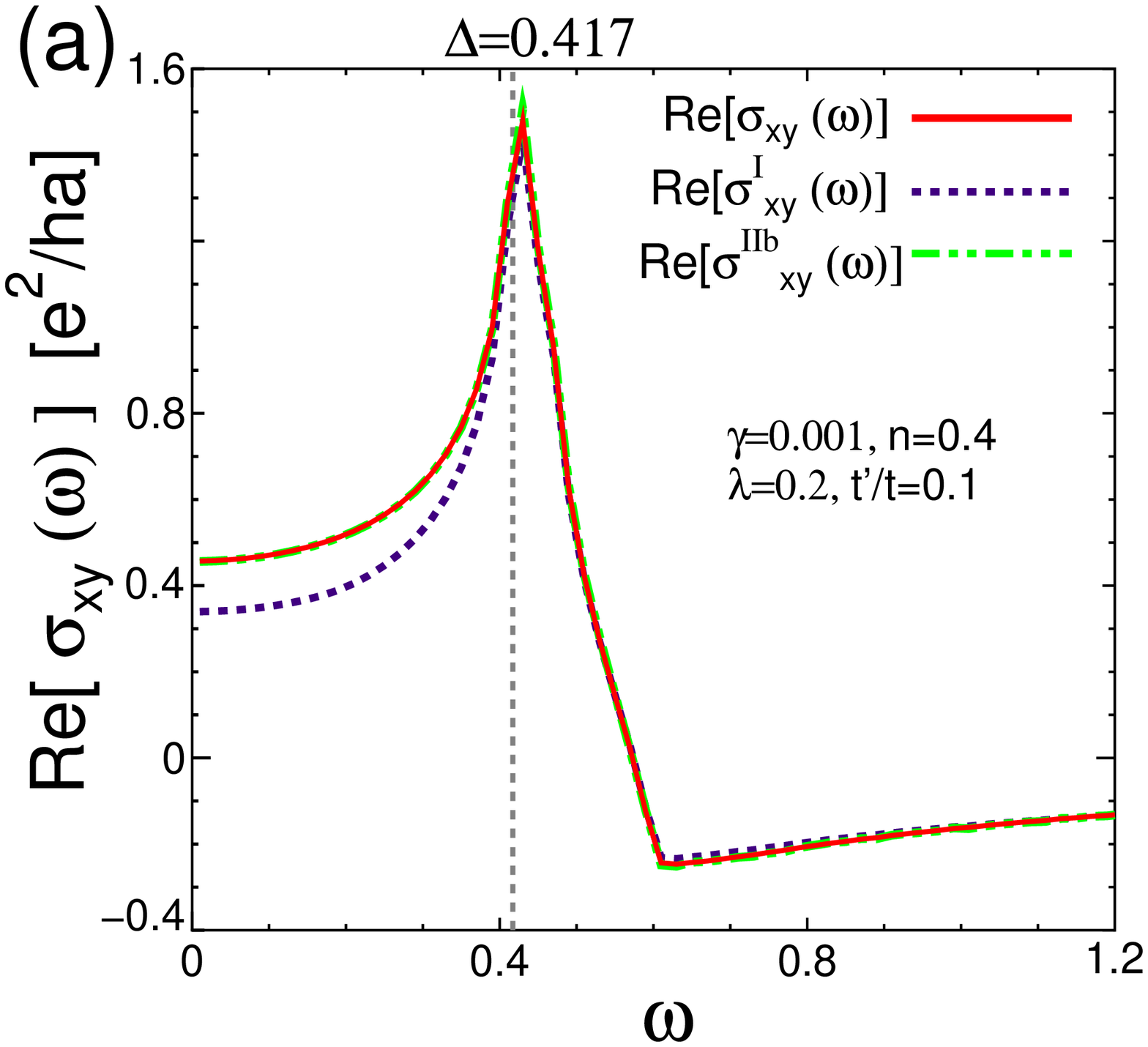}
\includegraphics[width=.85\linewidth]{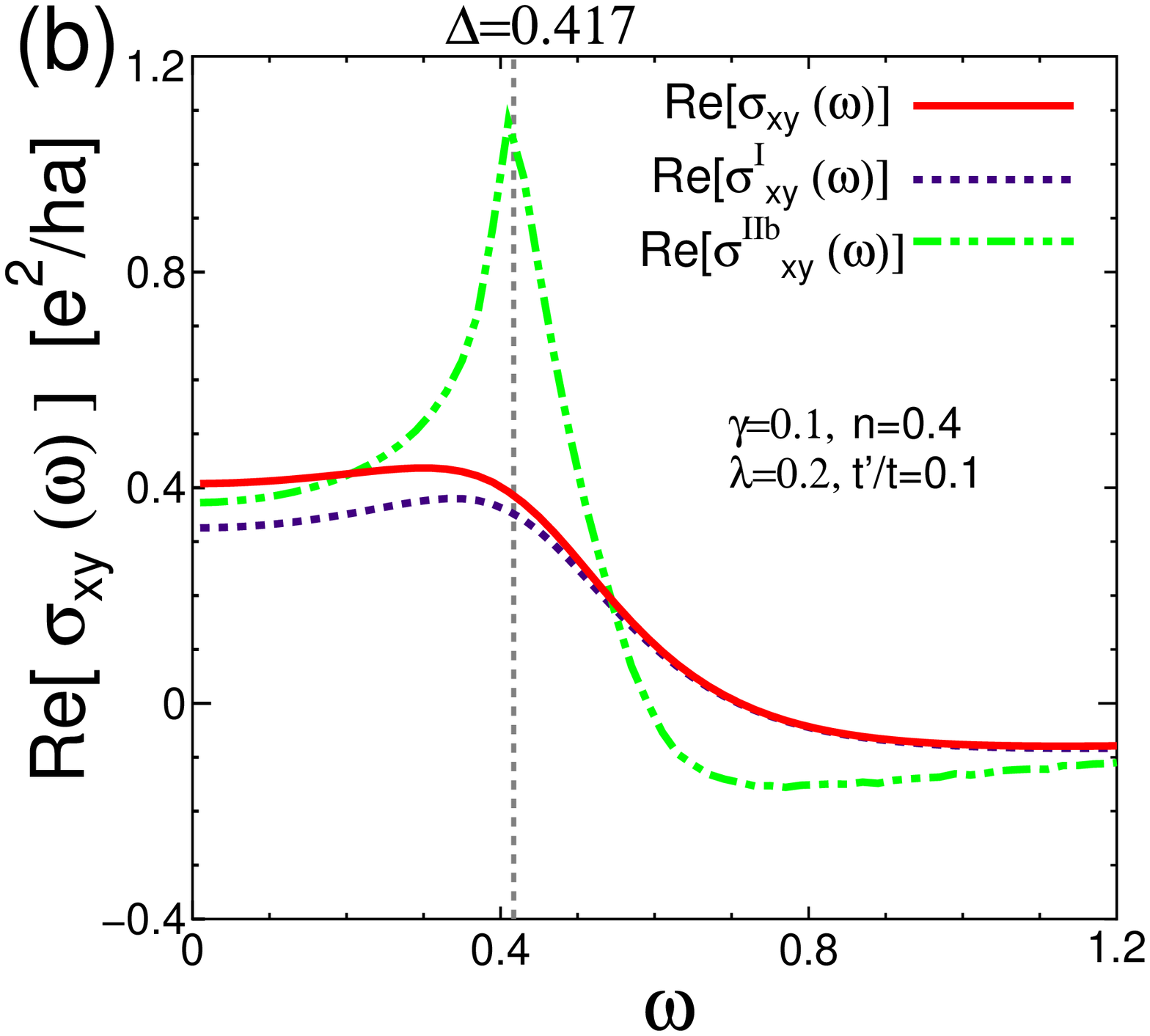}
\includegraphics[width=.85\linewidth]{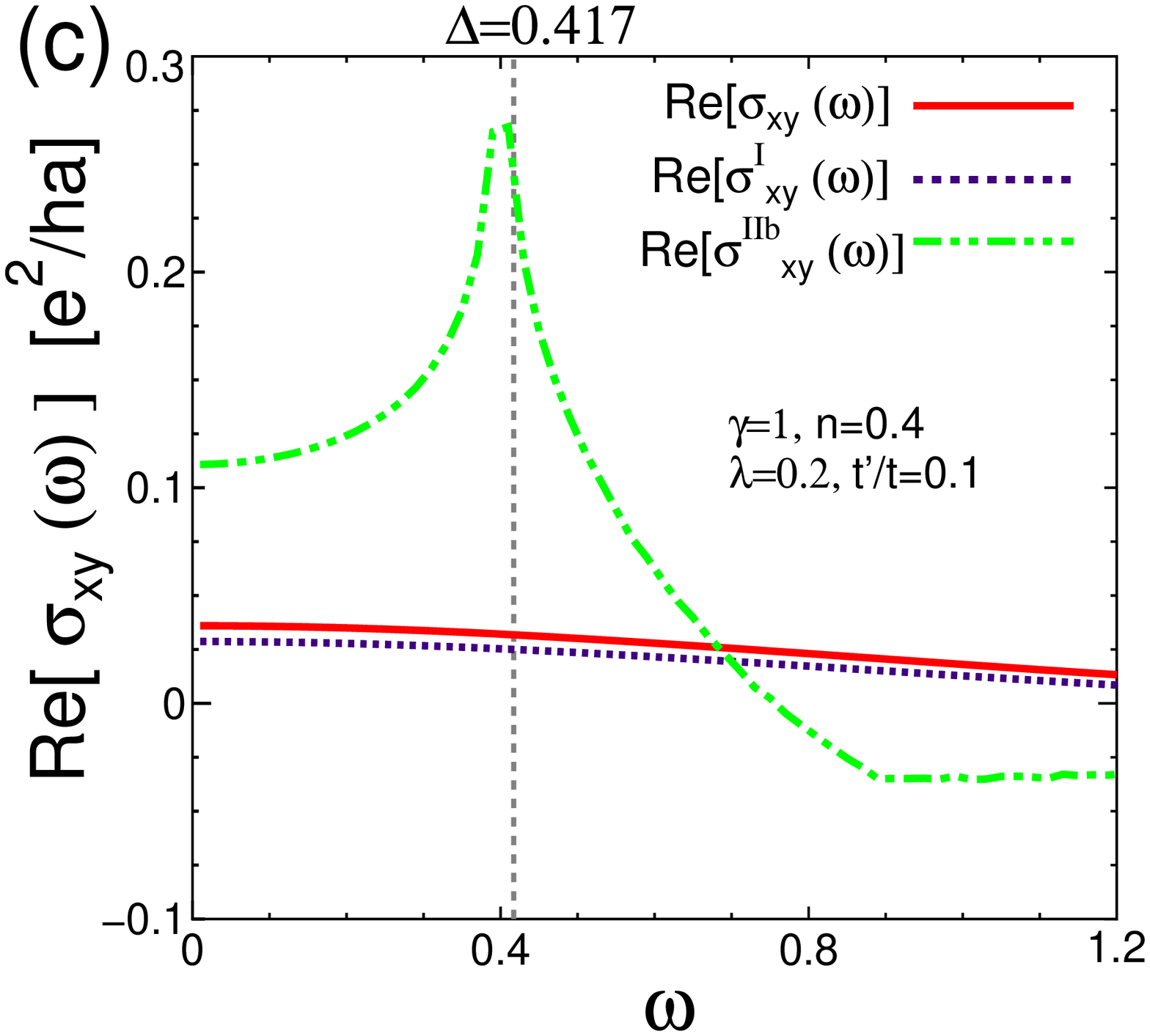}
\caption{\label{fig:pt1} $\w$-depenence of Re $\sxy(\w)$, Re $\sxy^{I}(\w)$ and Re $\sxy^{IIb}(\w)$ for $\g=0.001$ in (a), for $\g=0.1$ in (b), and for $\g=1$ in (c).
In (b) and (c), a sharp peak at $\w\sim\Delta$ in Re $\sxy^{IIb}(\w)$ is erroneous.}
\end{figure}

Figure \ref{fig:pt1} shows the $\w$-dependence of $\text{Re} \ \sxy(\w)$ for $\g=0.001, 0.1$, and 1.
When $\g=0.001$, $\text{Re} \ \sxy(\w)$ has a sharp peak at $\w\sim\Delta$ as shown in Fig. \ref{fig:pt1} (a). In this case, $\text{Re} \ \sxy^{IIb}(\w)$ (Berry curvature term) reproduces the total AHC $\text{Re} \ \sxy(\w)$ well.
On the other hand, we see from Fig. \ref{fig:pt1} (b) and (c) that the spiky peak of $\text{Re} \ \sxy(\w)$ is significantly suppressed when $\g$ is large. 
This fact is well reproduced by the Fermi surface term $\sxy^{I}(\w)$, whereas a sharp peak remains in the Berry curvature term $\sxy^{IIb}(\w)$ against large $\g$.
Therefore, $\sxy^{IIb}(\w)$ gives incorrect reslut of AHC when $\g$ is finite. 
We note that another Fermi sea term $\sxy^{IIa}(\w)$ is not shown in these figures since two Fermi sea terms satisfy the relationship $\sxy^{IIa}(\w) \sim -\sxy^{IIb}(\w)$ for any value of $\g$. 
As a result, the Fermi sea term is small in magnitude, and $\sxy^I(\w)$ gives a main contribution to $\sxy(\w)$. 
Thus, $\sxy^I(\w)$ repoduces the total AC AHC well for wide range of $\w$: 
\begin{eqnarray}
\sxy(\w) \sim \sxy^I(\w) \ \ \ (\text{Fermi surface term}).
\end{eqnarray}
This is one of the most important results in the present paper.
Recently, the LDA calculations of AHC and SHC in real systems were performed by
considering finite $\g$ \cite{Yao-GaAs, Xiao, Yao-CCSB}. However, the effect of $\g$ 
is underestimated in their calculations since only $\sxy^{IIb}$ is calculated.  

We have checked the reliability of the numerical study in two ways:
First, we verified that the AC conductivities (all $\sxx(\w), \sxy(\w), \sxy^{I}(\w)$, $\sxy^{IIa}(\w)$ and $\sxy^{IIb}(\w)$) reproduces the DC conductivities ($\w=0$) derived in ref. \cite{Kontani06}. 
Second, we calculated the sum rule for $\sxy(\w)$ numerically. The sum rule for $\sxy(\w)$ is given by

\begin{align}
\int^{\infty}_{0}d\w \text{Re} \ \sxy(\w) =0. \label{sumrule}
\end{align} 
Equation (\ref{sumrule}) is easily recognized from facts that $\sxy(\w)\sim|\w|^{-2}$ as $|\w|\rightarrow \infty$, and it is analytic in the upper-half plane of the complex $\w$-space. Here, we performed the numerical $\w$-integration of 
$r\equiv\int_{0}^{X}d\w \text{Re} \ \sxy(\w)/\int_{0}^{X}d\w|\text{Re} \ \sxy(\w)|$, where we put $X$=100. 
It should vanish identically when $X=\infty$ according to the sum rule. We have verified that $r\sim10^{-3}$, which suggests the high reliability of the present numerical study.  

\begin{figure}[!htb]
\includegraphics[width=.85\linewidth]{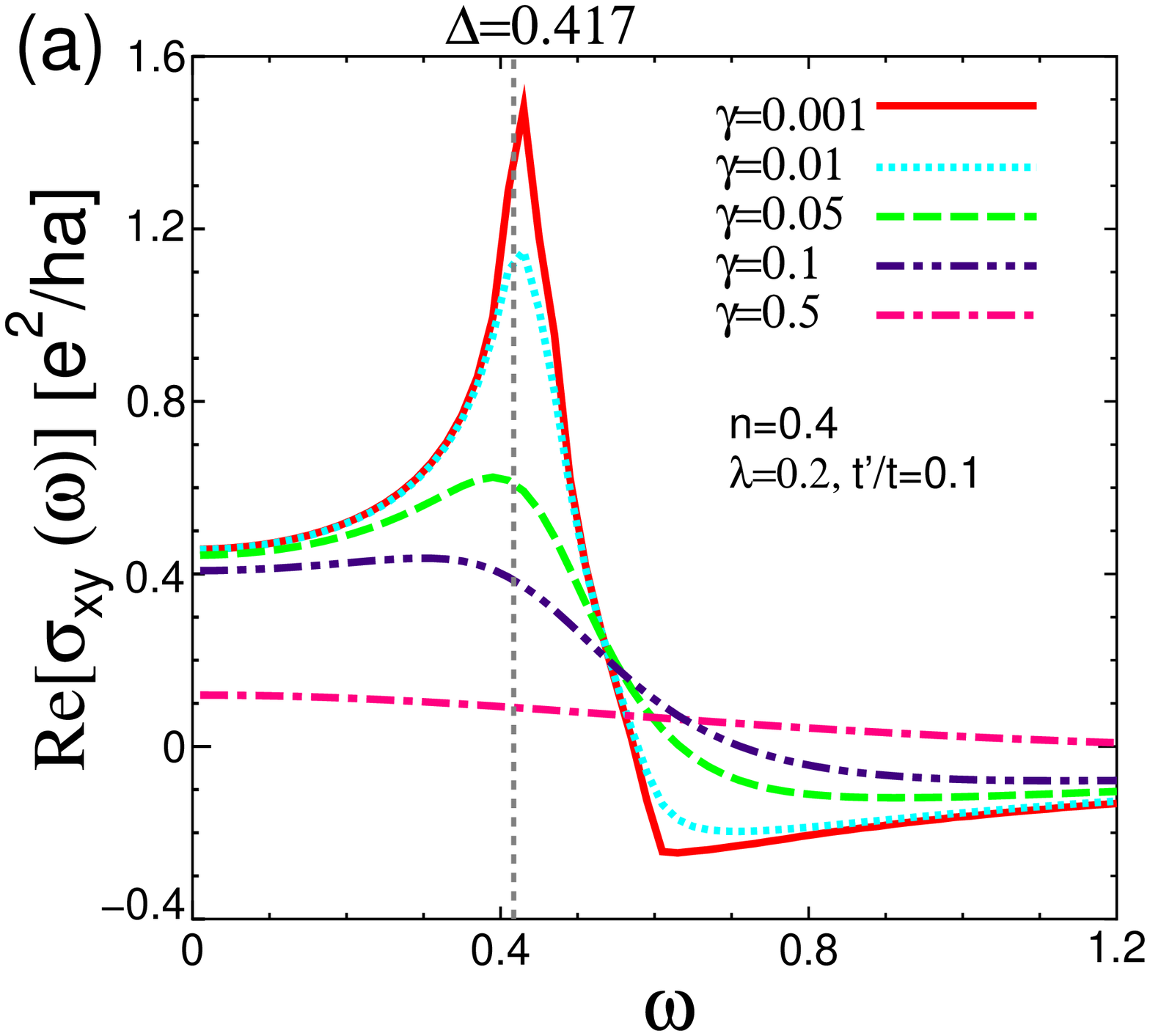}
\includegraphics[width=.85\linewidth]{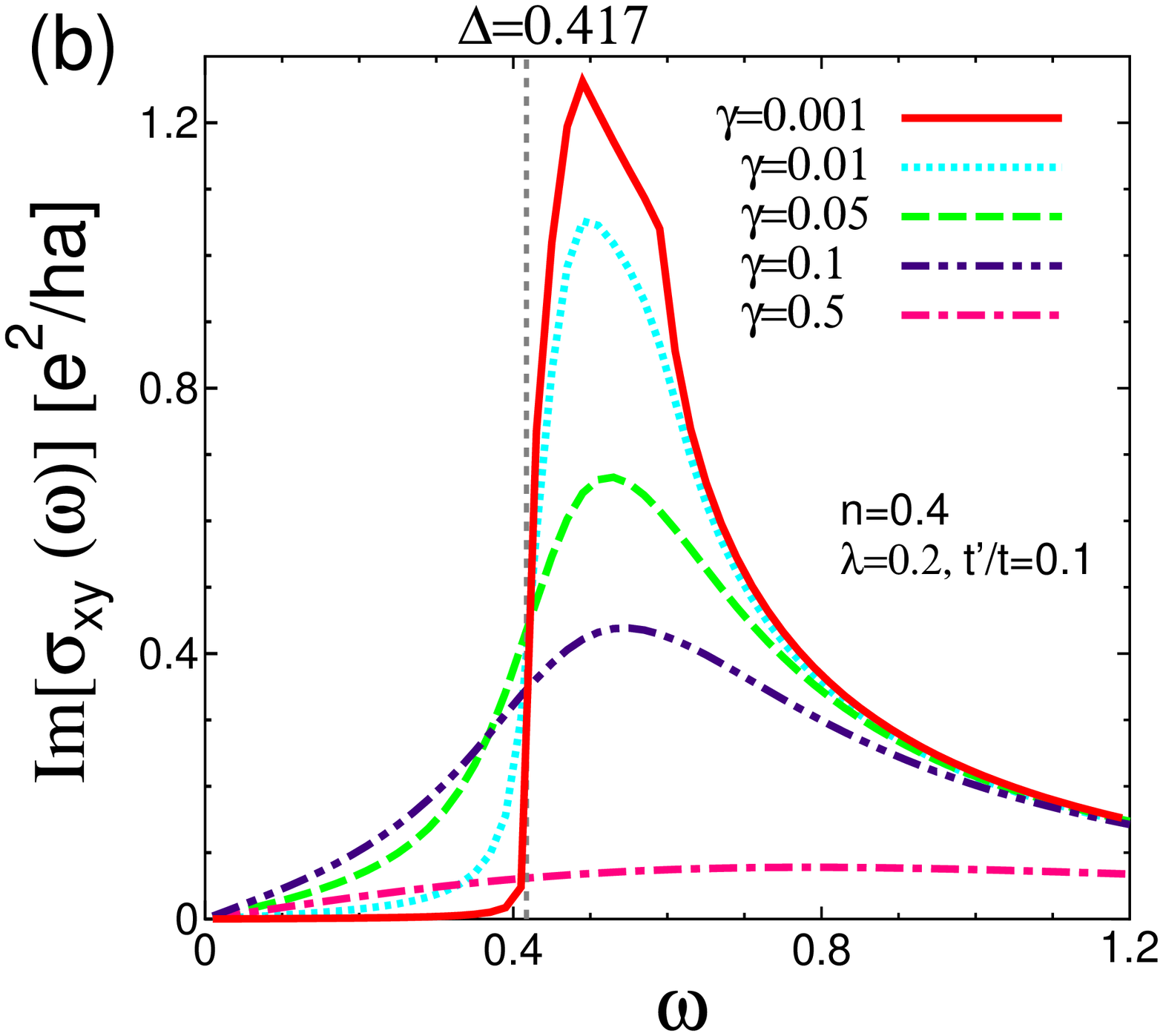}
\caption{ \label{fig:sxyomega} $\w$-dependence of $\sigma_{xy}(\omega)$ for various $\gamma$.
The real part of $\sxy(\w)$ is shown in (a), and the imaginary part in (b).
The spiky peak in Re $\sxy(\w)$ at $\w\sim\Delta$ is quickly suppressed by $\g$.}
\end{figure}

Hereafter, we show the numerical results of AC AHC in more detail. We first discuss the $\g$-dependence of the AC Hall conductivity $\sxy(\w)$.  
$\text{Re} \ \sxy(\w) $ for various values of $\g$ is shown in Fig. \ref{fig:sxyomega} (a). 
When $\g$ is very small, $\text{Re} \ \sxy(\w) $ has a sharp peak at finite energy $\Delta$. After reaching the peak at $\w\sim\Delta$, $\text{Re} \ \sxy(\w)$ decreases drastically and changes its sign. According to eqs. (\ref{eq:sxyw}), (\ref{eq:sxyIIbw}) and (\ref{eq:sxyIIaw}), the main contribution for $\text{Re} \ \sxy(\w)$ comes from area near $k^{\ast}$ in Fig. \ref{FS}. 
When $\g$ becomes large, however, $\text{Re} \ \sxy(\w)$ becomes almost constant 
for $0\leq \w \lsim \Delta$, and the peak at $\w\sim \Delta$ vanishes. 
In Fig. \ref{fig:sxyomega} (b), we show the $\w$-dependence of $\text{Im} \ \sxy(\w)$ for various $\g$. 
For $\g=0.001$ and $0.01$, $\text{Im} \ \sxy(\w)$ starts to increase drastically around $\w=\Delta$, and it takes a maximum value at $\w\sim 0.5$ which is slightly larger than $\Delta$.
We see that this peak is suppressed as $\g$ increases.

Now, we discuss the difference of the $\g$-dependences between DC AHC and $\text{Re} \ \sxy(\w)$. It is a well known property that DC AHC $\sigma_{xy}$ is independent of $\g$ in the low resistive regime where $\g\ll \Delta$ \cite{KL,Kontani06,Kontani94}. This property can be recognized in Fig. \ref{fig:sxyomega} (a) at $\w=0$. 
In contrast, at finite frequencies, Re $\sxy(\w)$ for $\g=0.001,0.01$, and 0.05 in Fig. \ref{fig:sxyomega} (a) behaves quite different from each other, especially around $\w\sim\Delta$.
Thus, the $\g$-dependence of the AC AHC is much more sensitive to the value of $\g$ compared with
the DC AHC. 



\begin{figure}[!htb]
\includegraphics[width=.85\linewidth]{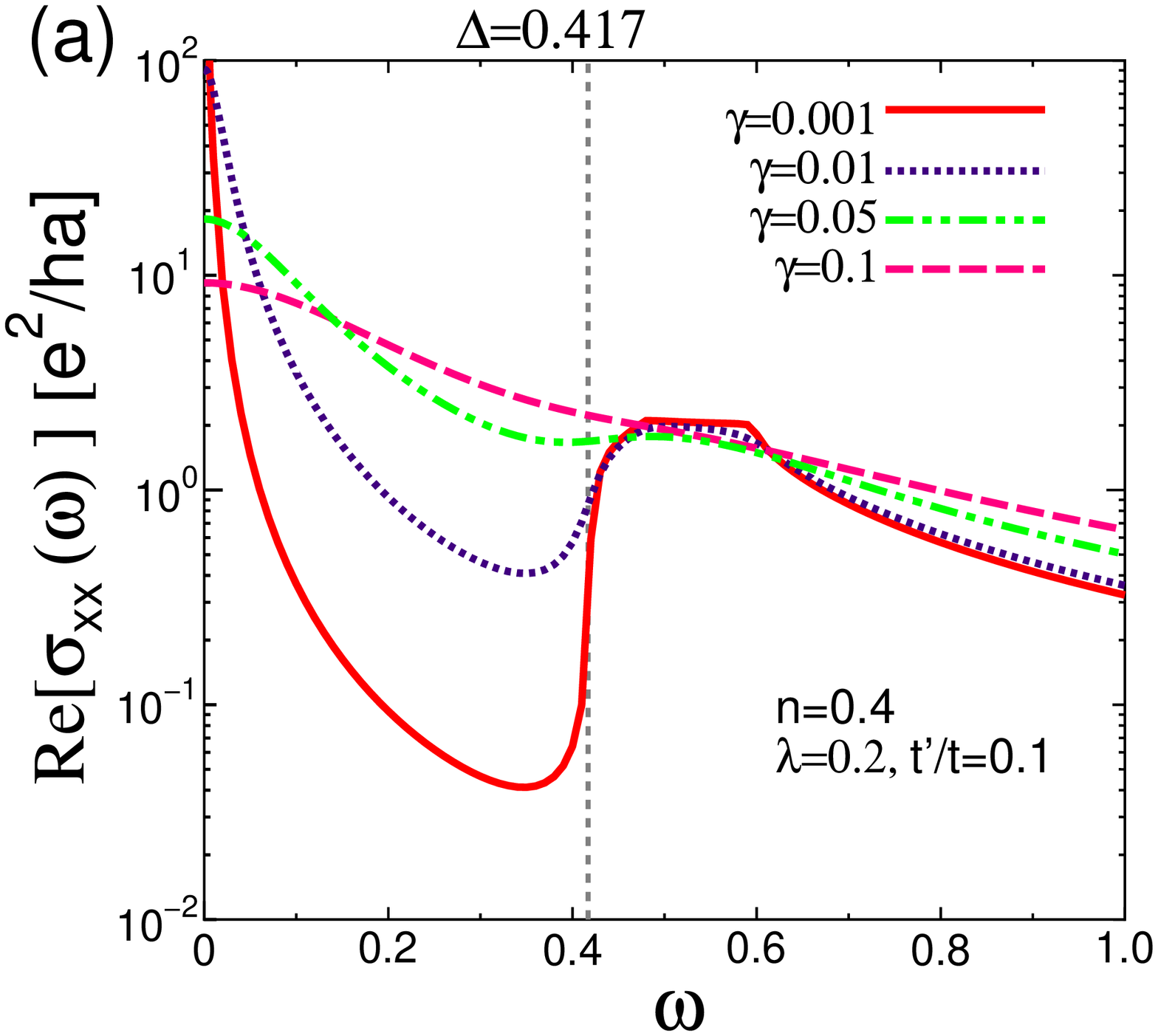}
\caption{ \label{fig:sxxomega} $\omega$-dependence of $\text{Re} \ \sigma_{xx}(\omega)$ for $\gamma=0.001, 0.01, 0.05, 0.1$ and 0.5.}
\end{figure} 

We also discuss the $\w$-dependence of $\text{Re} \ \sxx(\w)$.
Figure \ref{fig:sxxomega} shows that $\text{Re} \ \sxx$ has the Drude peak at $\w=0$.
It also shows a shoulder-type peak at $\w\sim\Delta$, which originates from the interband transition.

\begin{figure}[!htb]
\includegraphics[width=.85\linewidth]{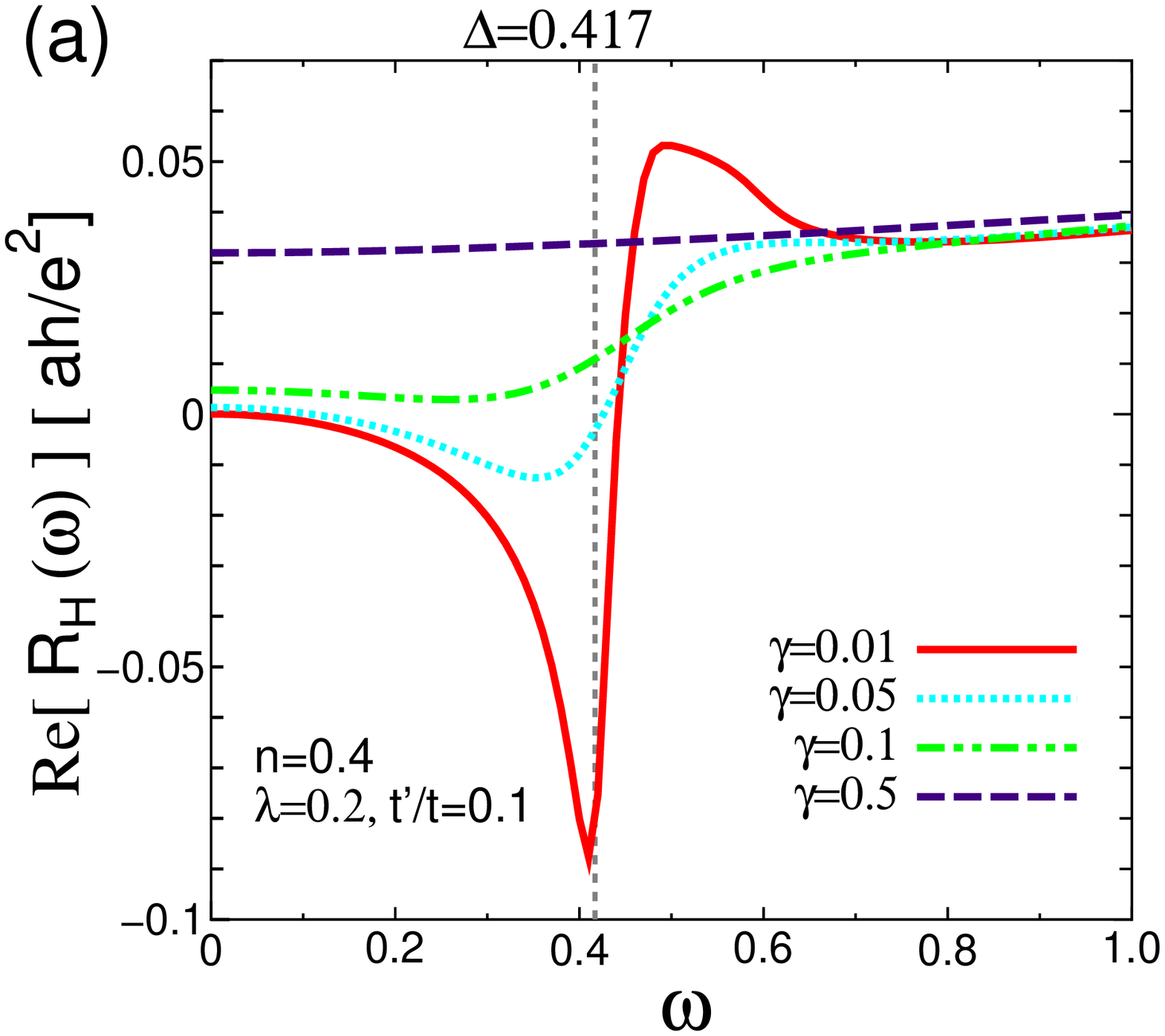}
\includegraphics[width=.85\linewidth]{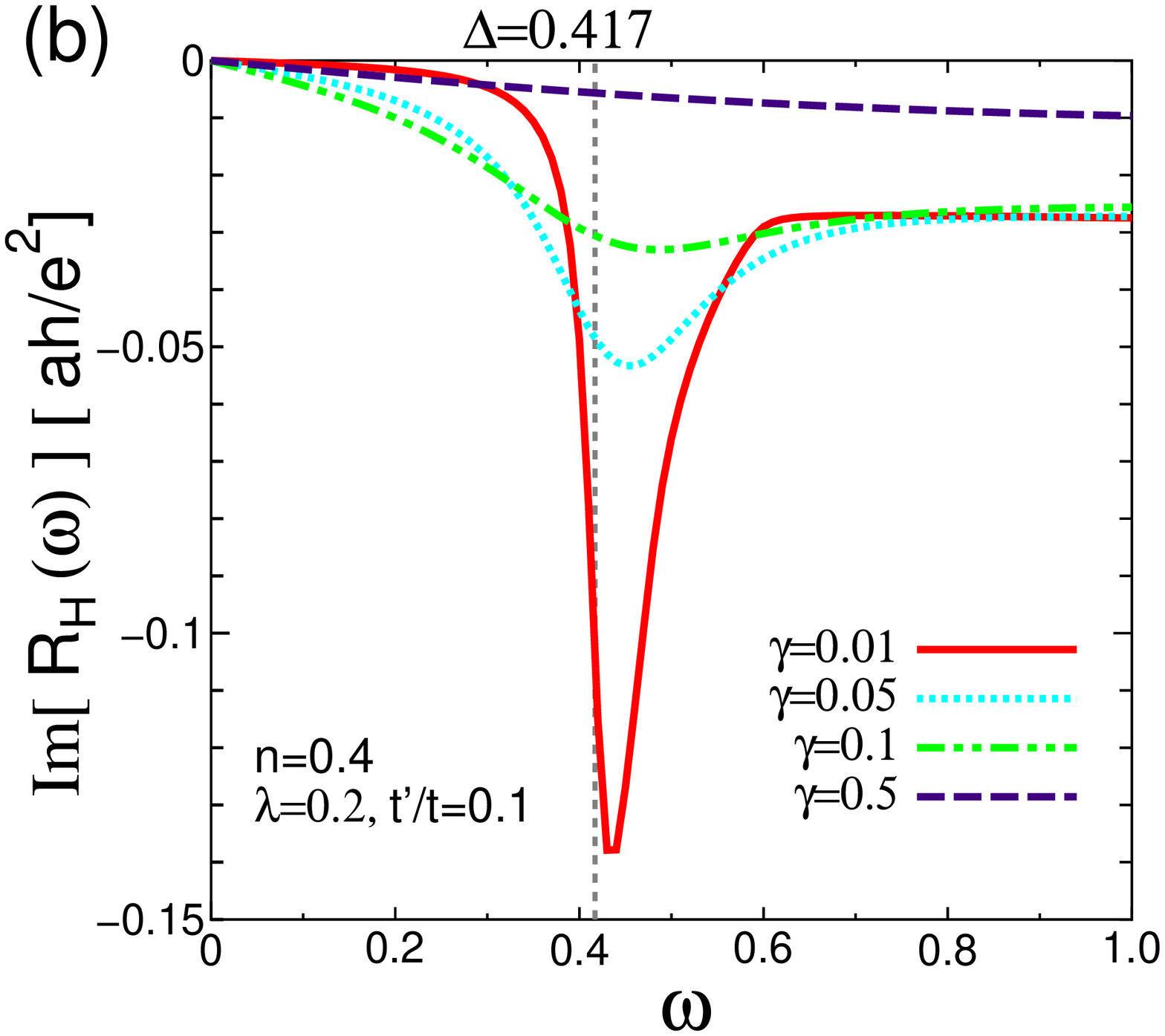}
\caption{ \label{fig:RH} $\w$-dependence of $R_H(\w)$ for $\gamma
= 0.01, 0.05, 0.1$ and 0.5.}
\end{figure}

\begin{figure}[!htb]
\includegraphics[width=.85\linewidth]{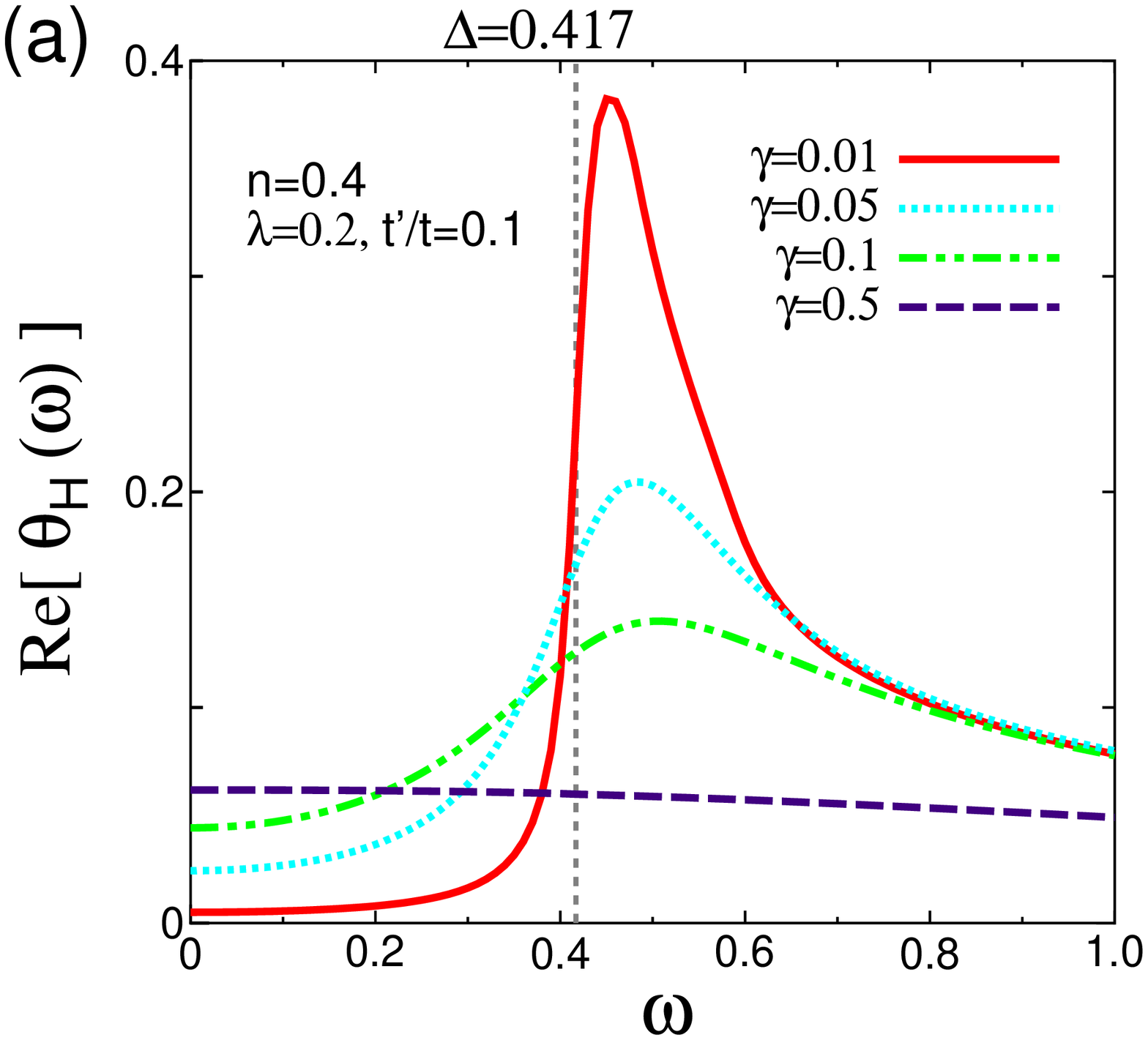}
\includegraphics[width=.85\linewidth]{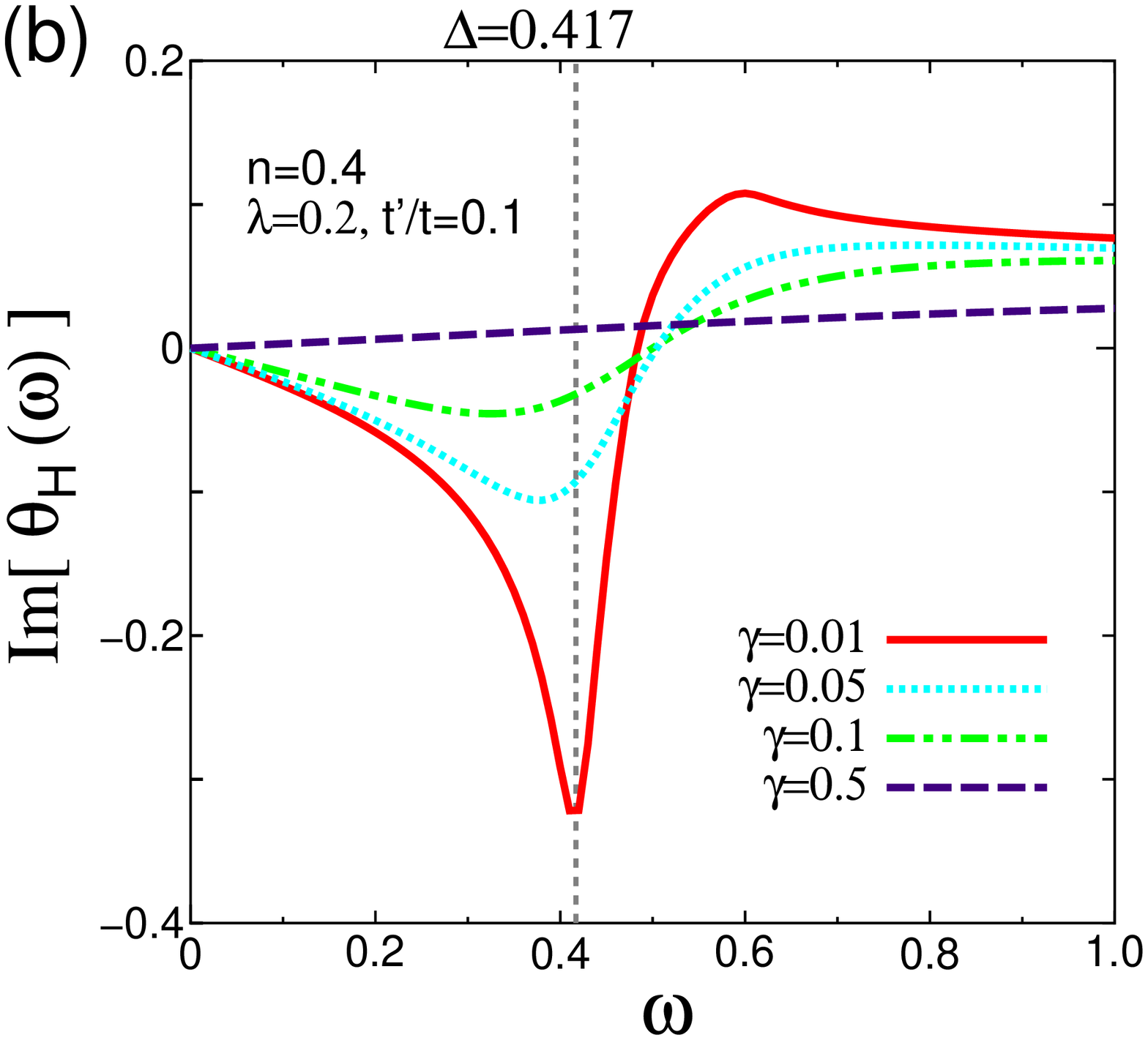}
\caption{ \label{fig:tH} $\w$-dependence of $\theta_H(\omega)$ for $\gamma
=0.01, 0.05, 0.1$ and 0.5.}
\end{figure}

Here, we examine the $\w$-dependence of the Hall coefficient $R_H(\w)=\sxy(\w)/(\sxx(\w))^2$ and the Hall angle $\theta_H(\w)=\sxy(\w)/\sxx(\w)$.
The real and imaginary part of $R_H(\w)$ is shown in Figs. \ref{fig:RH} (a) and (b), respectively.
From these figures, we see that both Re $R_H(\w)$ and Im $R_H(\w)$ changes significantly at $\w\sim\Delta$ for small $\g$. 
After reaching a peak at $\w\sim\Delta$, Re $R_H(\w)$ changes its sign, whereas Im $R_H(\w)$ remains negative.
However, this sharp peak at $\w\sim\Delta$ can be easily suppressed as $\g$ increases,
and Re $R_H(\w)$ for $\g=0.1$ and 0.5 remains negative.
As for the Hall angle, its $\w$-dependence is shown in Fig. \ref{fig:tH}.
When $\g$ is small, Im $\theta_H(\w)$ shows a peak at $\w\sim\Delta$, and it changes its sign for $\w>\Delta$.

Here, we briefly discuss the $\w$-dependence of $R_{H}(\w)$ and $\theta_{H}(\w)$ based on the simple Drude model. In the case of $\w\ll\Delta$, $\sxy^a(\w)\sim \sxy^a(0)$ and $\displaystyle \sxx(\w) \sim \f{\sxx(0)}{1-i\w\tau}$, where $\tau=1/2\g$.
Then, $R_{H}(\w) \sim R_{H}(0) (1-i\w\tau)^2 \propto (4\g^2 -\w^2) -4i\g\w$, and $\theta_{H}(\w)\sim \theta_{H}(0) (1-i\w\tau) \propto 2\g-i\w$. This analysis well explain the numerical results in Figs. \ref{fig:RH} and \ref{fig:tH} for $\w\ll\Delta$.


%
%

\section{\label{sec:level5} CALCULATION OF AC HALL CONDUCTIVITY WHEN $\gamma$ IS ENERGY DEPENDENT}

In the previous section, 
we calculated the AC AHC in the constant $\g$ approximation, assuming that the inelastic scattering due to local impurities are dominant.
However, in usual metals, inelastic scattering due to electron-electron interaction will be dominant, since the quasiparticle damping rate increases with $\w$:
In a Fermi liquid, the imaginary part of self-energy $\text{Im}\Sigma(\w)$ is given by
\begin{align}
\g(\w)=-\text{Im}\Sigma(\w)=\beta \ldk\lk \pi T\rk^{2} +  \w^2 \rdk + \g(0), \label{FLTgamma}
\end{align}
when $T$ and $\w$ are not large,
where $\beta$ is a constant and $T$ represents the temperature. $\g(0)$ represents the damping rate at the Fermi level due to elastic scattering. 
Here, we study the AC AHE when the quasiparticle damping rate $\g(\w)$ is given by eq. (\ref{FLTgamma}) by putting $\g(0)=0.005$.

When the damping rate $\g(\w)$ depends on $\w$, we cannot use eqs. (\ref{eq:sxyw}) - (\ref{eq:sxyIIbw}). Therefore, we perform the numerical calculations in eq. (\ref{basesxy}) for $\sxy(\w)$.
To perform this, we decompose eq. (\ref{basesxy}) in the difference of integral interval as follows:
\begin{align}
\sxy(\w)=& \sxy^A(\w) + \sxy^B(\w), \\
\sxy^A(\w)=&-\sum_{\bk,\alpha,\alpha',\beta,\beta'}\int^{\w/2}_{-\w/2}\f{d\eps}{2\pi\w}v^{\alpha'\alpha}_xv^{\beta'\beta}_y \nn
&\times G^R_{\alpha\beta'}(\eps+\w/2)
\ltk G^R_{\beta\alpha'}(\eps-\w/2)-G^A_{\beta\alpha'}(\eps-\w/2) \rtk,  \label{kukanw/2} \\
\sxy^B(\w)=&-\sum_{\bk,\alpha,\alpha',\beta,\beta'}\int^{-\w/2}_{-\infty}\f{d\eps}{2\pi\w}v^{\alpha'\alpha}_xv^{\beta'\beta}_y \nn 
&\times\ltk G^R_{\alpha\beta'}(\eps+\w/2)G^R_{\beta\alpha'}(\eps-\w/2)
- \langle R \leftrightarrow A \rangle\rtk. \label{kukaninft}
\end{align}
Here, we remind the readers that only the terms $(\alpha',\alpha,\beta',\beta)=(x,x,x,y),(x,x,y,x),(x,y,x,x)$ and $(y,x,x,x)$ remain finite after $\bk$-summation. Therefore, eqs. (\ref{kukanw/2}) and (\ref{kukaninft}) are rewritten as follows:  
\begin{align}
\sxy^A(\w)=&\sum_{\bk}\f{i\ld}{\pi\w}\int^{\w/2}_{-\w/2}d\eps v^{xx}_{x}v^{xy}_{y}  \nn 
         & \times\ldk \f{\w+i\g(\eps+\w/2)-i\g(\eps-\w/2)}{d^R(\eps+\w/2)d^R(\eps-\w/2)} \right. \nn
&-\left.\f{\w+i\g(\eps+\w/2)+i\g(\eps-\w/2)}{d^R(\eps+\w/2)d^A(\eps-\w/2)} \rdk,  
\label{kukan1w/2} \\
\sxy^B(\w)=&-\f{2\ld}{\pi\w}\sum_{\bk}\int^{-\w/2}_{-\infty}d\eps v^{xx}_{x}v^{xy}_y \nn
         &\times\ltk \w\times\text{Im}\ldk \f{1}{d^R(\eps+\w/2) d^R(\eps-\w/2)}\rdk \right. \nn 
&+ \ltk \g(\eps+\w/2)-\g(\eps-\w/2) \rtk \nn
&\left.\times \text{Re}\ldk \f{1}{d^R(\eps+\w/2 ) d^R(\eps-\w/2 )}\rdk \rtk. \label{kukan1inft}
\end{align}

We perform the $\bk$-summations in eqs. (\ref{kukan1w/2}) and (\ref{kukan1inft}) numerically, deviding
the Brilliouin zone into 1000 $\times$ 1000 meshes.
As for the numerical $\eps$-integration, we perform $\int^{-\w/2}_{X} d\eps$ in eq. (\ref{kukan1inft}), by setting $X=500$. Since the domain of $\eps$-integration in eq. (\ref{kukan1inft}) is about 100 times larger
than that in eq. (\ref{kukan1w/2}), $\eps$-integration is performed by dividing it into 50000 meshes for the former integration, and 500 meshes for the latter integration. 

\begin{figure}
\includegraphics[width=.85\linewidth]{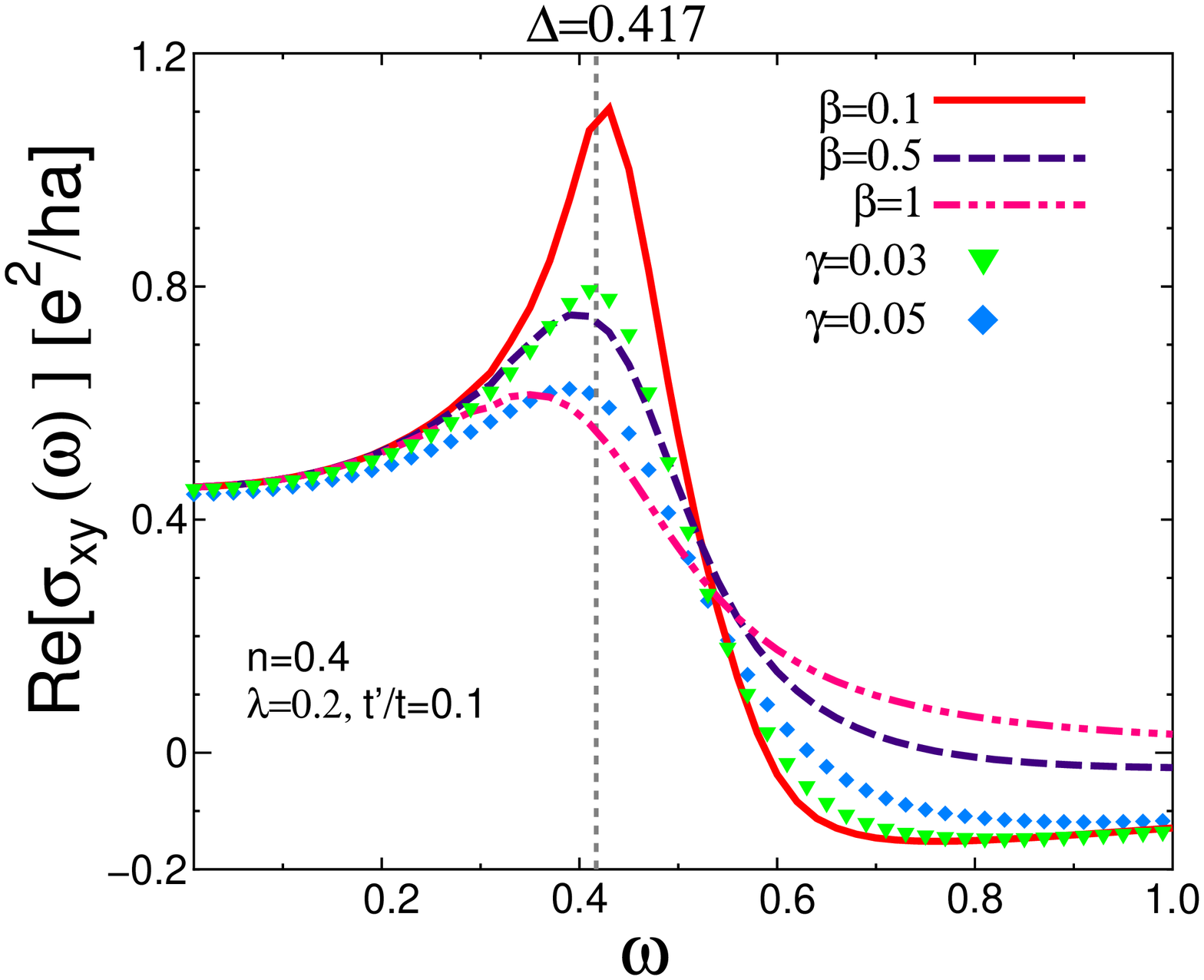}
\caption{\label{fig:sxybg} $\w$-dependence of $\text{Re} \ \sxy(\w)$ ($\beta=0.1,0.5$ and 1) when the dammping rate $\g(\w)$ is given by eq. (\ref{FLTgamma}). Here, $\text{Re} \ \sxy(\w)$ obtained for constant $\g$ ($\g= 0.03,0.05$) are shown for comparison.}
\end{figure}

Now, we show numerical results. Figure \ref{fig:sxybg} shows the $\w$-dependence of the AC Hall conductivity for $\beta= 0.1, 0.5$ and 1, which corresponds to $\g(\Delta/2)=0.0094,0.027$ and 0.049. 
The result of $\text{Re} \ \sxy(\w)$ for constant $\g$ ($\g=0.03$ and 0.05) are shown for comparison. 
We see that obtained Re $\sxy(\w)$ using eq. (\ref{FLTgamma}) 
is well reproduced by the constant $\g$ approximation for $\w\sim \Delta$, by putting $\g=\g(\Delta/2)$. 
To explain this fact analytically, we derive an analytical expression for $\text{Re} \ \sxy(\w)$ when $\g(\w)$ is $\w$-dependent. At zero temperature, $\sxy(\w)$ is mainly given by,
\begin{align}
\sxy(\w) &\propto \int^{0}_{-\w} d\eps \sum_{\bk}\theta(E^{+}_{\bk}-\mu)\theta(\mu-E^{-}_{\bk})  \nn
&\times \ldk \f{1}{\eps+\w+\mu-E^{+}_{\bk}+i\g(\eps+\w)}\f{1}{\eps+\mu-E^{-}_{\bk}-i\g(\eps)} \right.\nn
&+\left. \f{1}{\eps+\w+\mu-E^{-}_{\bk}+i\g(\eps+\w)}\f{1}{\eps+\mu-E^{+}_{\bk}-i\g(\eps)} \rdk. \label{eq:sxyge}
\end{align}  
We note that the first term represents the interband transition at $\w\sim E^{+}_{\bk}-E^{-}_{\bk}$, and the second term corresponds to $-\w\sim E^{+}_{\bk}-E^{-}_{\bk}$.  
Hereafter, we neglect the second term since we study $\w>0$ in the present study. By taking the pole of first term in eq. (\ref{eq:sxyge}), $\sxy(\w)$ can be estimated as
\begin{align}
\sxy(\w) &\propto i\int^{0}_{-\w} d\eps \sum_{\bk} \theta(E^{+}_{\bk}-\mu)\theta(\mu-E^{-}_{\bk})\nn
&\times\ldk \f{-\pi \delta (\eps+\w+\mu-E^{+}_{\bk})}{E^{+}_{\bk}-E^{-}_{\bk}-\w-i\g(\eps+\w)-i\g(\eps)} \right. \nn
 &+\left. \f{\pi \delta (\eps+\mu-E^{-}_{\bk})}{-E^{+}_{\bk}+E^{-}_{\bk}+\w+i\g(\eps+\w)+i\g(\eps)} \rdk.
\end{align}
Therefore, $\sxy(\w)$ for $\w\sim E^{+}_{\bk}-E^{-}_{\bk}$ is given as
\begin{align}
\sxy(\w) &\propto i\sum_{\bk} \f{\theta(E^{+}_{\bk}-\mu)\theta(\mu-E^{-}_{\bk})}{\w- ( E^{+}_{\bk}-E^{-}_{\bk})+i\g(E^{+}_{\bk}-\mu)+i\g(E^{-}_{\bk}-\mu)} . \label{eq:sxyge2}
\end{align}
In the case of $\w\sim\Delta$, the $\bk$-summation in eq. (\ref{eq:sxyge2}) is restricted around $\bk^{\ast}_{\pm}$.
Then, we approximate as $E^{+}_{\bk}\sim \mu + \Delta/2$ and $E^{-}_{\bk}\sim \mu -\Delta/2$ \cite{Kontani-RHletter,Kontani-RHfull}. 
As a result, we obtain the following extended-Drude (ED) expression for $\w\sim \Delta$:
\begin{align} 
\sxy(\w) \propto \f{i}{\w-\Delta+2i\g_{\text{ED}}(\w)},
\end{align}
where $\g_{\text{ED}}(\w)$ is approximately given by
\begin{align}
\g_{\text{ED}}(\w)&=\f{1}{2} [\g(\Delta/2)+ \g(-\Delta/2)] \nn
         &=\g(\Delta/2),
\end{align}
For example, when $\beta=1$ and $\g(0)=0.005$, the damping rate at $\w\sim\Delta$ is estimated as $\g(\Delta/2)\sim 0.049$. 
We also verified that $\text{Re} \ \sxy(\w)$ is mainly given by the Fermi surface term when the damping rate $\g(\w)$ depends on $\w$.
 
Here, we remark that eq. (\ref{FLTgamma}) is appropriate only for small $\w$. 
According to the perturbation thoery with respect to Coulomb interaction,
$\g(\w)$ will increase drastically for $|\w|>\Delta$ due to the interband excitation.
This fact will suppress $\sxy(\w)$ for $\w>2\Delta$ further.
Therefore, for a more reliable study, we have to calculate the $\w$-dependence of $\g(\w)$ microscopically.

Finally, we comment on two important future problems.
The first one is the detailed study of the Coulomb interaction effect on the AC AHE.
According to the microscopic Fermi liquid theory, the effect of Coulomb interaction is exactly renormalized to the self-energy correction and the CVC. 
As we have shown, the imaginary part of the self-energy tends to suppress the AC AHC.
As discussed in ref. \cite{Kontani06}, the renormalization factor due to the real part of the self-energy, 
$z=\left( 1- \left. \f{\partial \Sigma(\w)}{\partial \w} \right|_{\w=0} \right)^{-1}$, exactly cancels in the 
formula of the AC AHC given by eq. (\ref{eq:sxyw}). 
On the other hand, it is well-known fact that the CVC due to Coulomb interaction causes various anomalous transport 
phenomena in the vicinity of the magnetic quantum critical points (QCP)
\cite{Kontani-Hall,Kontani-MR,Kontani-S,Kontani-Nernst,Kontani-Yamada}.
This fact suggest that the CVC may
cause novel temperature dependence of the AC AHC near the magnetic QCP.
This is an important future problem.

Another future problem is to perform a more reliable calculation on AC AHC
based on a realistic tight-binding (TB) model. Recently, we have calculated SHC in 4$d$- and 5$d$-transition metals based on the Naval Research Laboratory 
tight-binding (NRL-TB) model \cite{Kontani-Pt, Tanaka-4d5d}. This model 
enables us to construct nine- orbital ($s+p+d$) TB models for each transition metal \cite{NRL1,NRL2}. In the future, we will study AC AHC based on 
this model to obtain more reliable results for AC AHC.


%
%

\section{\label{sec:level6} Discussions}
\subsection{\label{subsec:level6-1} Comments on Experiments}

In sections \ref{sec:level4} and \ref{sec:level5}, we have discussed the $\g$-dependence of AC AHC. We found that the spiky peak at $\Delta$ exists when $\g$ is very small, whereas it is easily suppressed as $\g$ increases. 
The value of the damping rate $\g(\w)$ at $\w\sim\Delta/2$ determines the characteristic behavior of the AC AHC. Here, we show that the spiky peak may vanish in $d$-electron systems by using the experimental value of $\g(\w)$: 
\v{C}erne et al. \cite{Drew} reported the damping rate $\g(\w)$ in Au and Cu. As for Cu (Au), the damping rate $\g(\w)$ is obtained as $\sim 600 \ (700) \ \text{cm}^{-1}$ when the frequency is $\sim 900 \ (1000) \ \text{cm}^{-1}$. 
This means that $\g\approx 600 \ \text{K}\sim 0.15$ at $\w\approx 900 \ \text{K} \sim 0.25$ in the present unit of energy. Since $\Delta \gsim 1000$K in usual transition metals \cite{Kontani-Pt,Tanaka-4d5d}, the observed damping rate is large enough to suppress the spiky peak of the AC Hall conductivity. 
Furthermore, the damping rate in the strongly correlated systems will be larger than those in Au and Cu. Therefore, we conclude that the peak of $\sxy(\w)$ at $\w\sim\Delta$ will be tiny or absent in usual transition metal ferromagnets.

As shown in Figs. \ref{fig:pt1} and \ref{fig:sxyomega}, the intrinsic AC AHC shows prominent deviation from the Drude-like behavior.
On the other hand, the AC AHC due to the skew-scattering mechanism is expected to follow the Drude-like behavior, $\sxy\propto (\g-i\w)^{-1}$ for small $\w$.
Therefore, AC AHC measurements will be quite useful to distinguish between the AHE due to intrinsic effect and that due to extrinsic effect, without necessity to introduce disorders.

%
%

\subsection{\label{subsec:level6-2} Summary of the Present Study}

In this paper, we studied the intrinsic AC AHE in transition metal ferromagnets based on $(d_{xz},d_{yz})$-orbital tight-binding model.
We drived an analytical expression for the AC AHC that is valid for any quasiparticle damping rate $\g$ without CVC, by performing the analytic continuation carefully.
We find that the intrinsic AC AHC does not follow the Drude-like behavior. When $\g$ is very small, AC AHC has a spiky peak at $\w\sim\Delta$, which arises from the interband transition as explained in Fig. \ref{fig:pt1} (a).
This behavior corresponds to the previously reported results by Fang et al. \cite{Fang}. 
When $\g$ is finite, however, the spiky peak is easily suppressed to be small or absent. In this case, the magnitude of AC AHC remains almost unchanged in the region $0< \w \lsim \Delta$.
We also calculated $\sxx(\w)$, the Hall coefficient $R_{H}(\w)$, and the Hall angle $\theta_H(\w)$:
$R_{H}(\w)$ and $\theta_H(\w)$ show a sharp peak at $\w=\Delta$ for small $\g$, whereas this peak is easily suppressed as $\g$ increases as shown in Figs. \ref{fig:RH} and \ref{fig:tH}.

The overall behavior of the AC AHC $\sxy(\w)$ is reproduced by the Fermi surface term $(I)$, as in the case with the DC AHC. The Fermi surface term strongly depends on $\g$, whereas the Berry curvature term $(IIb)$ has a weak dependence of $\g$. Although the relation $\sxy(\w) \approx \sxy^{IIb}(\w)$ holds in the present model when $\g$ is very small,
the relation $\sxy(\w)\approx \sxy^{I}(\w)$ is well satisfied for a wide range of $\g$. 
Therefore, prominent $\g$-dependence of AC AHC is well reproduced by the Fermi surface term.
We stress that $\sxy(\w)\neq\sxy^{IIb}(\w)$ even if $\g=+0$ in general multiorbital systems \cite{Kontani06,KontaniSHE}.
For a quantitative study of the intrinsic AHC, however, we have to calculate both the Fermi surface term and the Fermi sea terms on the same footing. 

Finally, we comment on the AC SHE.
Recently Kontani et al. \cite{KontaniSHE} have studied the intrinsic SHE in $d$-electron systems.
Therein, they have found that the present ($d_{xz},d_{yz}$) tight-binding model shows huge SHE. 
In the present model, SHC $\sxy^{z}$ is given by $(-\hbar/e)$ times the AHC $\sxy$: 
$\sxy^z = -\f{\hbar}{e}\sxy$, since the spin of the conduction electron is conserved. 
Therefore, interesting $\w$-dependence of AC AHC derived in the present study 
is also expected to be realized in AC SHC in various transition metal complexes.

\begin{acknowledgments}

We are grateful to D.S. Hirashima, K. Yamada, J. Inoue and Y. Suzumura for fruitful discussions.
This study has been supported by Grants-in-Aid for Scientific
Research from the Ministry of Education, Culture,
Sports, Science and Technology of Japan.
Numerical calculation were performed at the facilities
of the Supercomputer Center, ISSP, University of Tokyo.

\end{acknowledgments}

\end{document}